\documentclass[amsmath,superscriptaddress,longbibliography,prl,twocolumn] {revtex4-2}

\usepackage{dsfont}
\usepackage{amsmath}
\usepackage{epsfig}
\usepackage{graphicx}
\usepackage{bm}
\usepackage{amssymb}
\usepackage{slashed}
\usepackage{hyperref}
\hypersetup{
     colorlinks   = true,
     citecolor    = blue,
     urlcolor = blue,
     linkcolor = blue
}

\begin{document}
% ==============================================================================

\title{Anisotropic resistivity and superconducting instability in ferroelectric metals}

\author{Vladimir~A.~Zyuzin}
\affiliation{L.~D.~Landau Institute for Theoretical Physics, Semenova~1-a,~142432, Chernogolovka, Russia}

\author{Alexander~A.~Zyuzin}
\affiliation{QTF Centre of Excellence, Department of Applied Physics, Aalto University, FI-00076 AALTO, Finland}

% ==============================================================================
\begin{abstract}
We propose a theoretical model of a ferroelectric metal where spontaneous electric polarization coexists with the conducting electrons. In our model we adopt a scenario when conducting electrons interact with two soft transverse optical phonons, generalize it to the case when there is a spontaneous ferroelectric polarization in the system, and show that a linear coupling to the phonons emerges as a result. We find that this coupling results in anisotropic electric transport which has a transverse to the current voltage drop. Importantly, the obtained transverse component of the resistivity has distinct linear dependence with temperature.
Moreover, we show that the coupling enhances superconducting transition temperature of the ferroelectric metal. 
We argue that our results help to explain recent experiments on ferroelectric strontium titanate, as well as provide new experimental signatures to look for.   
\end{abstract}
\maketitle

The strontium titanate-based (STO) compound has a rich phase diagram upon chemical doping and temperature variation, which includes seemingly self-exclusive ferroelectric and superconducting states, see for a review 
\cite{Fernandes_review, Behnia_review, Muller_review}. To begin with, we note that the pristine $\mathrm{SrTiO}_3$ is a wide-band gap quantum paraelectric insulator. It can be tuned into a ferroelectric state by partial 
substitution of Sr ions with Ca, Ba, or Pb \cite{Bednorz_Ca, Lemanov_Ba, Lemanov_Pb}, isotope substitution of oxygen \cite{Itoh_para_ferro_isotop}, and applying stress \cite{Stress_para_ferr_trans}.
The structural formation is associated with the soft transverse optical (TO) lattice vibration. 
The gap in the dispersion of this phonon mode vanishes at the transition \cite{Fernandes_review, Behnia_review}.

On the other hand, STO becomes semiconducting (or metallic) with partial substitution of Sr with Nb, La, or with oxygen reduction, see for details \cite{Fernandes_review, Behnia_review, Muller_review}.
The charge transport measurements show the square temperature dependence of resistivity within an unusually wide region of material parameters, \cite{Behnia_review, Behnia_T2_review}. 
Another remarkable property of the material is the existence of the superconductivity despite of the rather low electron density.
Studies of superconductivity in this material have a very long history, \cite{Fernandes_review, Behnia_review, Muller_review}, dating back to experimental work \cite{Schooley_first_experiment}. 
However, the mechanism of Cooper pairing in STO is still currently under debate.
First of all, the superconducting transition temperature mediated by acoustic phonons in STO was estimated to be negligibly small \cite{Ruhman_Lee}, thus ruling out conventional pairing mechanism due to phonons.
Moreover, the superconductivity in this system emerges in a situation when Fermi energy is an order of magnitude smaller than the Debye frequency. 
Furthermore, the proximity of the system to a ferroelectric quantum critical point suggests that soft TO phonons might be important.

A possible solution to the problem of superconductivity in STO was proposed a long time ago by Ngai, who introduced 
a model of superconducting instability based on electron coupling tuned by a pair of TO phonons \cite{Ngai_main}.
In paraelectric STO the electron scattering by transverse phonons is proportional to the second power in lattice displacement amplitude. Below we will be referring to this mechanism as the two-phonon.
The two-phonon mechanism is distinct from the electron scattering by acoustic phonons, which is described by the gradient of lattice displacement. Later Epifanov, Levanyuk, and Levanyuk studied 
the T-squared dependence of conductivity due to the two-phonon mechanism near ferroelectric phase transition \cite{ELL_conductivity_1, ELL_conductivity_2}.

Recent observation of the enhancement of superconducting transition temperature in STO upon oxygen isotope substitution, which brings the system closer to the ferroelectric transition, 
has reinvigorated this subject, \cite{Paraelectric_superconductor_exp}. 
Proximity to a ferroelectric instability strongly suggests that the soft TO phonons play an influential role. 
Thus the model of two-phonon scattering was brought forward to revisit the temperature dependence of charge transport 
\cite{Maslov_conductivity, Feigelman_conductivity} and study the effect of oxygen isotope substitution on superconducting transition temperature \cite{Marel_2phonon, Kiseliov_Feigelman, Volkov_ferro_super}.

The story is far from being complete. 
Recently, the signatures of ferroelectric instability has been observed in n-doped $\mathrm{Sr}_{1-x}\mathrm{Ca}_x\mathrm{TiO}_{3-\delta}$, \cite{Ferroelectric_superconductor_exp, Ferroelectric_T_exp, Ferroelectric_superconductor_exp_2,Tomioka2022}.
It was found that superconductivity and ferroelectricity may coexist in this material \cite{Ferroelectric_superconductor_exp, Ferroelectric_superconductor_exp_2,Tomioka2022}. The resistivity showed anomalous temperature dependence, suggesting the emergence of an additional scattering channel, \cite{Ferroelectric_T_exp, Ferroelectric_superconductor_exp_2,Tomioka2022}. The coexistence of metallic and ferroelectric phases might be understood within the dipole model. The ferroelectric instability is based on the interplay between the long-range dipole-dipole interaction and the short-range repulsion between ions. The former favors ferroelectric structure formation with the emergence of electric dipole moment per unit cell and the latter supports the paraelectric phase, please see \cite{Cohen_review} and for a review \cite{Ferroelectric_review}. In metals, itinerant electrons screen dipole-dipole interaction and thus eliminate ferroelectricity. However, supported by the experiments \cite{Ferroelectric_superconductor_exp, Ferroelectric_T_exp, Ferroelectric_superconductor_exp_2,Tomioka2022}, ferroelectricity may survive in weakly doped semiconductors presumably due to the long Thomas-Fermi screening length or local-bonding contributions \cite{Benedek}. 

Motivated by these experiments, \cite{Ferroelectric_superconductor_exp, Ferroelectric_T_exp, Ferroelectric_superconductor_exp_2,Tomioka2022}, we propose and study a model of electron scattering by one-phonon in the presence of spontaneous ferroelectric polarization. The model is a generalization of the two-phonon mechanism \cite{Ngai_main} to the ferroelectricity, and adds up with two-phonon mechanism, \cite{Ngai_main, ELL_conductivity_1, ELL_conductivity_2}.  
We show that below the structural transition, the decrease of temperature enhances electric resistivity due to the onset of ferroelectric polarization. At lower temperatures, we predict linear in temperature suppression of resistivity.
We also calculate the contribution of one-phonon scattering processes to the superconducting transition temperature. We argue that our mechanism might dominate over the two-phonon mediated pairing deep in the ferroelectric phase.

%---------------------------------------------------------------------------------------------------------------------
\textit{Model.} 
%---------------------------------------------------------------------------------------------------------------------
We start with introducing a model of a ferroelectric metal (polar metal), which consists of electrons interacting with TO phonons.
The TO phonons are responsible for the ferroelectricity in the system, while the electrons are responsible for the electric conduction.
The electrons are described by the Hamiltonian
\begin{align}
H_{\mathrm{e}} = \int_{\bf r} \psi^{\dag}(x) \left( -\frac{\partial^2_{\bf r}}{2m } - \mu \right)\psi(x),
\end{align}
where
$\psi^\dag(x), \psi(x)$ are the electron creation and annihilation operators, $\mu$ is the chemical potential, $m$ is the mass of electrons, $x=({\bf r}, t)$ is a general coordinate, and $\int_{\bf r}(..) \equiv \int d{\bf r}(..)$.
The Hamiltonian of two degenerate denoted by $a=1,2$ branches of TO phonons is 
\begin{align}
H^{(a)}_{\mathrm{ph}} = \sum_{\bf q} \omega_{\bf q}\left(b_{a;{\bf q}}^{\dag}b_{a;{\bf q}} + \frac{1}{2}\right),
\end{align}
where
$
\omega_{\mathbf{q}} = \sqrt{\omega_{\mathrm{TO}}^2+s^2q^2}
$
is the spectrum with $s$ being the speed of sound and $\omega_{\mathrm{TO}}$ being the TO phonon gap. 
Polarization vector $\mathbf{P}(x)$ describing the TO phonons is
\begin{eqnarray}
\mathbf{P}(x)  = 
\sum_{a;\mathbf{q}}
\frac{\mathbf{e}_{a;\mathbf{q}}}{\sqrt{V}}A_{\mathbf{q}} 
\left[ 
b_{a;\mathbf{q}}(t)
e^{i\mathbf{q r}} 
+ 
b^{\dag}_{a;\mathbf{q}}(t)
e^{-i\mathbf{q r}}  \right] ,~~~
\end{eqnarray}
where $V$ is the volume of the material, $\mathbf{e}_{a;\mathbf{q}}$ is the unit vector in the direction of polarization of $a=1,2$ branches of TO phonons with wave-vector $\mathbf{q}$, and
$b^{\dag}_{a;\mathbf{q}},~ b_{a;\mathbf{q}}$ are the bosonic creation and annihilation operators. We consider 
$ A_{\mathbf{q}}^2  = \Omega_{0}^2/(4\pi \omega_{\bf q}),
$
where $\Omega_{0}$ is material dependent coefficient determined via Lyddane-Sachs-Teller relation for the static dielectric function
$\epsilon_0(\mathbf{q})=\Omega_0^2/\omega_{\mathbf{q}}^2$, \cite{Maslov_conductivity}. 
The identity unit vectors satisfy is 
\begin{align}\label{orth}
\sum_{a=1,2} e_{a;{\bf q},\alpha}e_{a;{\bf q}, \beta} = \delta_{\alpha\beta} - \frac{q_{\alpha}q_{\beta}}{q^2},
\end{align}
where $\alpha,\beta = x,y,z$. We assume that $\omega_{\mathrm{TO}}$ vanishes at temperature $T_{\mathrm{FE}}$ (ferroelectric transition temperature), and the system becomes ferroelectric below this temperature forming spatially homogeneous spontaneous electric polarization $\langle {\bf P}(x)\rangle = {\bf P}_{0}$. For temperatures $T<T_{\mathrm{FE}}$ the phonon gap is $\omega_{\mathrm{TO}}\propto \vert {\bf P}_{0}\vert$.

We assume that electrons interact with two TO phonons, \cite{Ngai_main, ELL_conductivity_1, ELL_conductivity_2}. Namely, it is impossible for electrons to interact with one TO phonon via $\propto \left[\mathrm{div}\cdot{\bf P}(x)\right]\psi^\dag(x)\psi(x)$ term (of the Frohlich type), however, it is possible for electrons to interact with two TO phonons via $\propto {\bf P}(x)\cdot {\bf P}(x)\psi^\dag(x)\psi(x)$ term. Below we will be referring to this mechanism as the two-phonon mechanism.
In what follows, we generalize the two-phonon mechanism to the ferroelectric case, when $\langle {\bf P}(x) \rangle = {\bf P}_{0}$ in the system. 
We then write for the interaction between electrons and the TO phonons,
\begin{equation}\label{main_Hamiltonian}
H_{\mathrm{e-ph}} = g\int_{\bf r} [ \mathbf{P}_{0} + \mathbf{P}(x)]^2\psi^\dag(x)\psi(x),
\end{equation}
where $g$ is the electron-phonon interaction constant. We will be using $\hbar = k_{\mathrm{B}} = 1$ units throughout the paper.

%%%%%%%%%%%%%%%%%%%%%%%%% FIGURE 1 %%%%%%%%%%%%%%%%%%%%%%%%%
\begin{figure}[t!]
\centering
\includegraphics[width=0.33\textwidth]{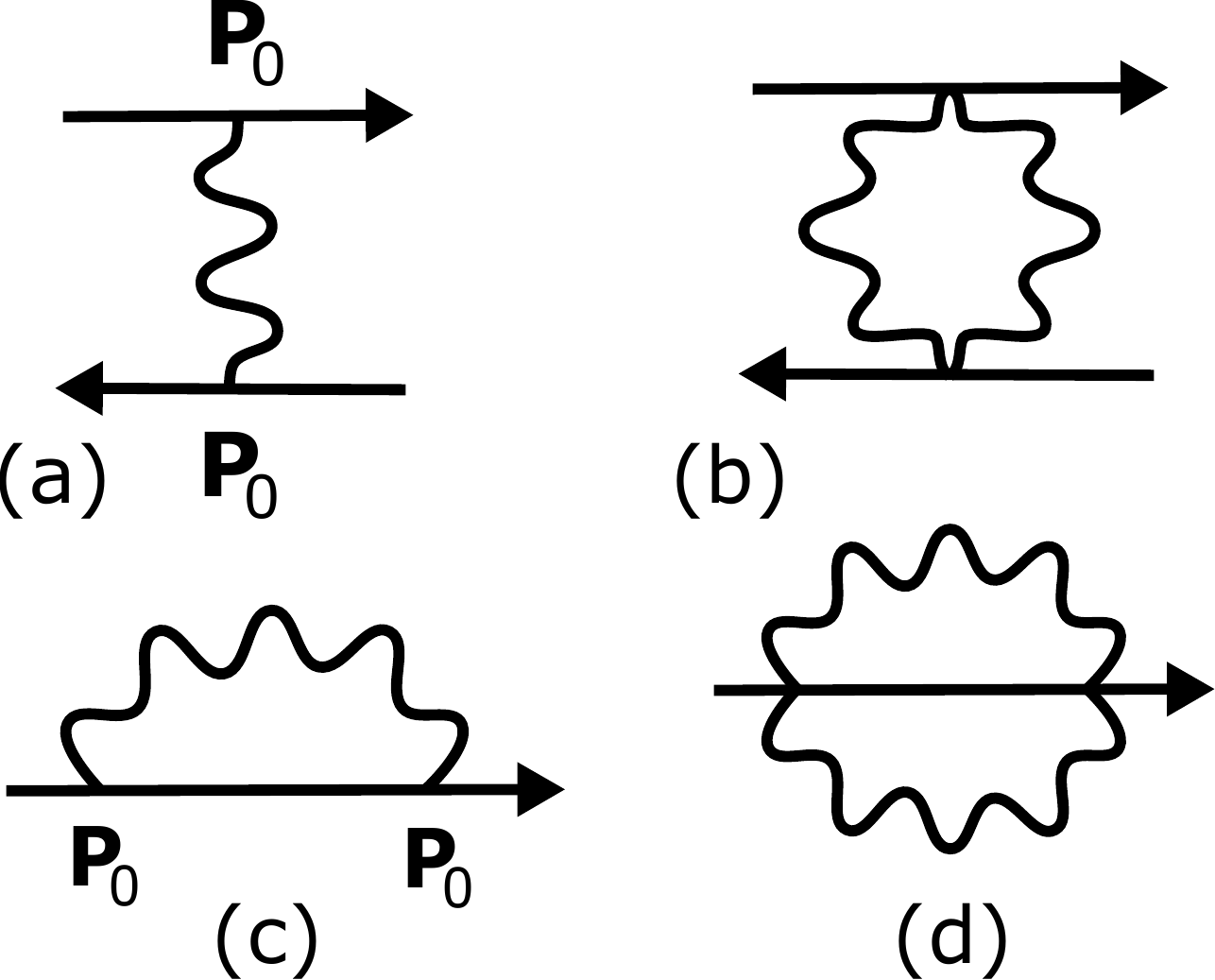}
\caption{\label{fig1} Feynman diagrams for (a,b) the effective electron interaction and (c,d) self-energy. Here the wavy line is the phonon Green function. Object ${\bf P}_{0}$ in (a,c) denotes the one-phonon contribution to the interaction in case of ferroelectric order. Figures (b,d) describe two-phonon interaction.}
\end{figure}
%%%%%%%%%%%%%%%%%%%%%%%%% FIGURE 1 %%%%%%%%%%%%%%%%%%%%%%%%%

We note in passing that we do not consider Coulomb repulsion between the electrons and focus on the electron-phonon interaction only.
The reason for that is the large dielectric constant. 
Moreover, we assume that the electrons don't screen finite electric polarization ${\bf P}_{0}$.

%---------------------------------------------------------------------------------------------------------------------
\textit{Effective electron interaction.} 
%---------------------------------------------------------------------------------------------------------------------
Here we discuss corrections to electron quantum life-time and conductivity due to the electron-phonon interaction.
We choose to work in Keldysh formalism as it conveniently describes fluctuations of the system about its equilibrium at finite temperatures, please see Supplemental Material 
\cite{SM_editors}.

In order to obtain effective electron interaction, as usual, we integrate out the phonons.
To second order in electron-phonon interaction we obtain two processes shown in Fig. (\ref{fig1}a) and (\ref{fig1}b).
Processes Fig. (\ref{fig1}b) describe interaction of electrons with two TO phonons [due to $\propto {\bf P}^2(x)$ in Eq. (\ref{main_Hamiltonian})], while those in Fig. (\ref{fig1}a) describe interaction of electrons with one TO phonon [due to $\propto {\bf P}_{0}\cdot{\bf P}(x)$ in Eq. (\ref{main_Hamiltonian})] given that there is a spontaneous electric polarization ${\bf P}_{0}$ in the system. 
The processes in Fig. (\ref{fig1}b) were studied in Refs. \cite{Ngai_main, ELL_conductivity_1, ELL_conductivity_2, Maslov_conductivity, Feigelman_conductivity, Marel_2phonon, Kiseliov_Feigelman, Volkov_ferro_super}, 
and we refer to these works for details, the processes in Fig. (\ref{fig1}a) are new and are subject of the following analysis. 
The effective electron interaction due to these processes [wavy line in Fig. (\ref{fig1}a) and (\ref{fig1}c)] is
\begin{align}\label{int1}
V_{1}^{\mathrm{R}/\mathrm{A}}({\bf q};\omega) = (2g P_{0} A_{\bf q})^2
 \sin^2(\phi_{{\bf q}{\bf P}_{0}})
 D^{\mathrm{R}/\mathrm{A}}({\bf q};\omega),
\end{align} 
where $D^{\mathrm{R}/\mathrm{A}}({\bf q};\omega) = 2\omega_{{\bf q}}/[(\omega \pm i0)^2 - \omega_{{\bf q}}^2]$ is the TO phonon (retarded/advanced) Green function. We have defined an angle between ${\bf q}$ and ${\bf P}_{0}$ as $\cos(\phi_{{\bf q}{\bf P}_{0}})\equiv ({\bf q}\cdot {\bf P}_{0})/(qP_{0})$. 

With all the details given in the SM \cite{SM_editors} we here present essential results and experimental predictions of the model.

%---------------------------------------------------------------------------------------------------------------------
\textit{Self-energy and resistivity.}
%---------------------------------------------------------------------------------------------------------------------
The experiments, see for a review Refs. \cite{Fernandes_review, Behnia_review} show that the resistivity of metallic STO is proportional to $T^2$ at low temperatures. 
In a typical Fermi liquid, such temperature dependence originates from the electron-electron Coulomb interaction.
However, in metallic STO, due to a large dielectric constant, Coulomb interaction is expected to be weak. 
In Refs. \cite{ELL_conductivity_1, ELL_conductivity_2, Maslov_conductivity, Feigelman_conductivity, Marel_2phonon} it was theoretically suggested that the $T^2$ contribution to the resistivity due to the two-phonon mechanism can be significant.
Namely, the self-energy Fig. (\ref{fig1}d) due to the electron-phonon processes which are depicted in Fig. (\ref{fig1}b) results in the decay-rate proportional to the $T^2$, which is independent of the electronic density of states \cite{Maslov_conductivity}.

In another set of experiments Refs. \cite{Ferroelectric_superconductor_exp, Ferroelectric_T_exp, Ferroelectric_superconductor_exp_2} it was observed that substitution of Sr atoms with Ca, i.e. by creating Sr$_{1-x}$Ca$_{x}$TiO$_{3}$ compound, results in the ferroelectric transition of the material. Moreover, it was shown that the ferroelectric order survives when the Sr$_{1-x}$Ca$_{x}$TiO$_{3}$ is made metallic by doping it. Furthermore, the experiments clearly observe a new boson mode the conducting electrons scatter by in the ferroelectric phase, which results in the non-monotoneous temperature dependence of the resistivity in the vicinity of the ferroelectric transition \cite{Ferroelectric_superconductor_exp, Ferroelectric_T_exp, Ferroelectric_superconductor_exp_2}.

%%%%%%%%%%%%%%%%%%%%%% BEGIN FIGURE %%%%%%%%%%%%%%%%%%%
\begin{figure}[t!]
\includegraphics[width=0.37\textwidth, height=0.3\textheight]{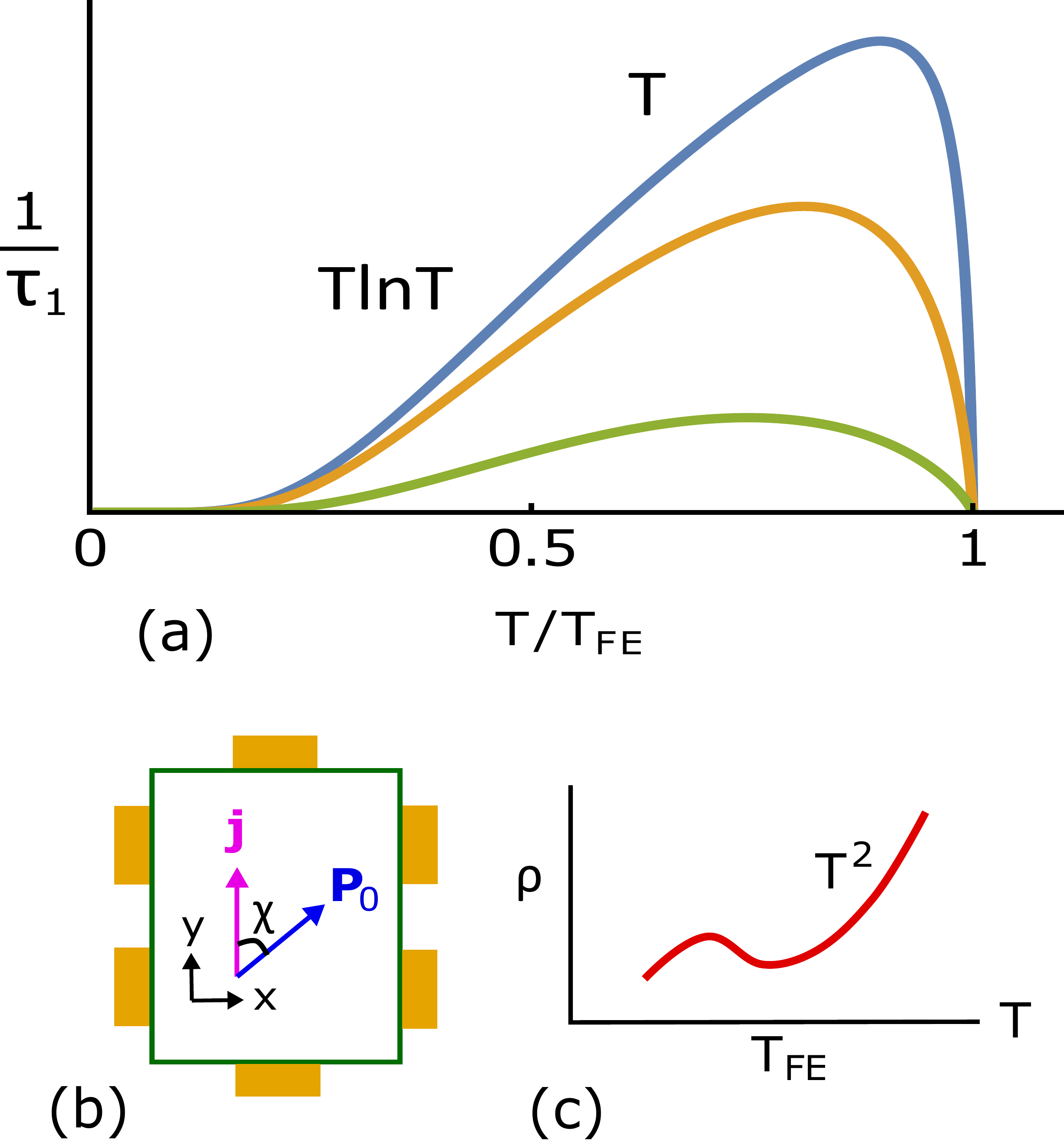}
\caption{\label{fig2} (Color online). (a) Temperature dependence of the decay rate due to electron scattering by one TO phonon. 
Assuming temperature dependence in $P_0^2(T) \propto \omega_{\mathrm{TO}}^2(T) = \omega_{\mathrm{TO}}^2(0)(1-T/T_{\mathrm{FE}}) $, the curves are plotted for $T_{\mathrm{BG}}/\omega_{\mathrm{TO}}(0) = \{0.2,0.5,1.5\}$. 
The amplitude decreases with the increase of $T_{\mathrm{BG}}/\omega_{\mathrm{TO}}(0)$.
(b) Schematics of the Hall bar to measure the electric transport anisotropy given in Eq. (\ref{result2}). 
With the direction of the electric polarization being in $x-y$ plane, the current is passed at some angle $\chi$ with respect to the polarization, for example, in $y-$direction.
(c) Schematics of the temperature dependence of the longitudinal resistivity given in Eq. (\ref{result2}). For $T>T_{\mathrm{FE}}$ the resistivity is expected to be $\propto T^2$ due to two-phonon mechanism, Refs. \cite{ELL_conductivity_1, ELL_conductivity_2, Maslov_conductivity, Feigelman_conductivity}.
 Below $T_{\mathrm{FE}}$ the one-phonon mechanism Eq. (\ref{result1}) depicted in Fig. (\ref{fig2})a sets in, in addition to two-phonon $\propto T^2$, resulting in an increase of the resistivity.
}
\end{figure}
%%%%%%%%%%%%%%%%%%%%%% END FIGURE %%%%%%%%%%%%%%%%%%%

Let us analyze how the electron scattering on one-phonon interaction derived in Eq. (\ref{int1}) affects the resistivity of the system.
Taking the imaginary part of the electron-self energy Fig. (\ref{fig1}c) and setting frequency to zero, we find the electron decay rate due to scattering via one-phonon mechanism, $\tau_{1{\bf k}}^{-1}$. 
Setting momentum of the electron $k = k_{\mathrm{F}} \equiv \sqrt{2m \mu}$, we obtain
\begin{eqnarray}\label{result1}
\frac{1}{\tau_{1{\bf k}}}  &=& 2\left[1+\cos^2(\phi_{{\bf k}{\bf P}_{0}}) \right] \left(2\pi g \frac{\Omega_0 P_0}{T_{\mathrm{BG}}} \right)^2 \nu \\
&\times& T \ln\left| \mathrm{tanh}\left(\frac{\sqrt{T^2_{\mathrm{BG}}+\omega_{\mathrm{TO}}^2}}{2T}\right)\mathrm{coth} \left(\frac{\omega_{\mathrm{TO}}}{2T}\right)\right|,
\nonumber
\end{eqnarray}
where $\nu = m k_{\mathrm{F}}/2\pi^2$ is the density of electron states per spin and $T_{\mathrm{BG}} = 2sk_{\mathrm{F}}$ is the Bloch-Gruneisen temperature.
Let us separate isotropic part from the angle-dependent one by a redefinition of $\tau_{1{\bf k}}^{-1} \equiv \tau_{1}^{-1}\left[1+\cos^2(\phi_{{\bf k}{\bf P}_{0}}) \right] $. 

The temperature dependence of the isotropic part of the decay rate is shown in Fig. (\ref{fig1}a).
At the ferroelectric transition $P_{0} \rightarrow 0$ in the prefactor of the Eq. (\ref{result1}) taking care of the logarithm, which formally diverges due to $\omega_{\mathrm{TO}} \rightarrow 0$ at the transition. Hence, right at the transition 
the decay rate vanishes as shown in Fig. (\ref{fig1}a).
Slightly below the ferroelectric transition, $T_{\mathrm{FE}}\gtrsim T$, the decay rate increases with the decrease of temperature as shown in Fig. (\ref{fig2}). We may assume $P_0$ and $\omega_\mathrm{TO}$ to be temperature independent far from the transition. In this case, at $T_{\mathrm{FE}}>T \gtrsim T_{\mathrm{BG}},\omega_{\mathrm{TO}}$, the decay rate scales linearly with temperature, 
and for the region $T_{\mathrm{FE}}, T_{\mathrm{BG}}>T>\omega_{\mathrm{TO}}$, we find $\tau_1^{-1} \propto T \ln |2T/\omega_{\mathrm{TO}}|$. 
As another example, let's note that $\tau_{1} \propto k_{\mathrm{F}}$ at $T>T_{\mathrm{BG}}$. It is drastically different from the two-phonon process, which is independent of $k_{\mathrm{F}}$ at these temperatures \cite{Maslov_conductivity}.
It is also instructive to compare the obtained decay rate with the one due to electron scattering on acoustic longitudinal phonons (obtained from the Frohlich type interaction). 
There, the decay rate is $\propto \nu T/T^2_{BG}$ at large temperatures $T\gg T_{\mathrm{BG}}$, while $\propto \nu T^3/T_{\mathrm{BG}}^2$ at low temperatures $T \ll T_{\mathrm{BG}}$, \cite{Allan, Migdal}.   

Let us calculate the electric current in the system. 
In the SM \cite{SM_editors}, we construct the kinetic equation with the collision integral defined by the impurity, one- and two-phonon scattering processes. The kinetic equation is solved to give the electric current
\begin{align}\label{result2}
{\bf j}
=
\sigma
\left(1 - \frac{\tau}{5\tau_{1}} \right){\bf E}
- \sigma \frac{2\tau}{5\tau_{1}} ({\bf E}\cdot{\bf P}_{0}){\bf P}_{0}\frac{1}{P_{0}^2} ,
\end{align}
in which $\sigma =  e^2\nu D$ is the conductivity, where $D=v_{\mathrm{F}}^2\tau/3$ is the diffusion coefficient, and 
$
\tau^{-1} =  \tau_{\mathrm{imp}}^{-1} + \tau_{1}^{-1}  + \tau_{2}^{-1} 
$
is the total decay rate due to the isotropic impurity, one- and two-phonon scattering processes (in general decay rate due to the electron-electron Coulomb interaction should also be included).

Obtained anisotropy of the electric current [second term in Eq. (\ref{result2})] results in the transverse responses. For example, in Fig. (\ref{fig2}b) we schematically show the Hall-bar setup, where electric current is passed in $y-$ direction at some angle $\chi$ to the electric polarization, which is in $x-y$ plane. The voltage drop in the direction perpendicular to the current is measured either in opposite or along diagonal gates to study the angle dependence, $\propto \cos(\chi)\sin(\chi)$, of the transverse part of Eq. (\ref{result2}). 
Most importantly, as can be deduced from Eq. (\ref{result2}), only the anisotropic part of the resistivity depends on temperature as shown in Fig. (\ref{fig2}a).

The longitudinal part of the resistivity in the ferroelectric phase is more complicated as the two-phonon mechanism (also impurity scattering and others) contributes to it as well.
In Fig. (\ref{fig2}c) we schematically plot the temperature dependence of the longitudinal part of the resistivity Eq. (\ref{result2}). 
There, for $T>T_{\mathrm{FE}}$ the resistivity is expected to be $\propto T^2$ due to two-phonon mechanism as predicted in Refs. \cite{ELL_conductivity_1, ELL_conductivity_2, Maslov_conductivity, Feigelman_conductivity}.
Below $T_{\mathrm{FE}}$ the one-phonon mechanism Eq. (\ref{result1}) sets in resulting in an increase of the resistivity with a characteristic dip at $T_{\mathrm{FE}}$.
The two-phonon mechanism exists in the ferroelectric phase as well because $\omega_{\mathrm{TO}}\rightarrow 0$ at the transition, allowing for the $\propto T^2$ results of Refs. \cite{ELL_conductivity_1, ELL_conductivity_2, Maslov_conductivity, Feigelman_conductivity} to be applicable. We argue that this picture explains experimental results of \cite{Ferroelectric_superconductor_exp, Ferroelectric_T_exp, Ferroelectric_superconductor_exp_2}.

We conclude that by subtracting from the longitudinal resistivity its temperature-squared part of the fit in the ferroelectric phase, one should explicitly obtain a one-phonon contribution with a characteristic temperature dependence shown in Fig. (\ref{fig2}a).
In addition, the same temperature dependence, shown in Fig. (\ref{fig2}a), is expected in the measured transverse voltage.

%---------------------------------------------------------------------------------------------------------------------
\textit{Superconducting transition temperature.}
%---------------------------------------------------------------------------------------------------------------------
Let us now proceed with the calculation of correction of one-phonon process to the Cooper pairing in ferroelectric metal. 
To qualitatively estimate the superconducting transition temperature, we adopt approach of Refs. \cite{Barkhudarov_Tc, Gorkov_Tc}.
The pole in the fermion scattering amplitude in the Cooper channel determines the superconducting transition temperature. Equation for the respective vertex part is given by
\begin{align}\label{SQE}
\Gamma({\bf q}) = V({\bf q}) 
- 
\int_{\bf p} V({\bf q}-{\bf p}) 
\frac{\mathrm{tanh}\left(\frac{\xi_{{\bf p} }}{2T} \right)  }{2\xi_{{\bf p}} } \Gamma({\bf p}).
\end{align}
where $\xi_{{\bf p}} = \mathbf{p}^2/2m -\mu$, $ V({\bf q}) = V_1({\bf q})+V_2({\bf q})$ is the electron-electron attraction potential to be specified for one and two TO phonon coupling mechanisms, $V_1({\bf q})$ and $V_2({\bf q})$ respectively, in what follows. 

Let us first revisit two-phonon mechanism of superconductivity in paraelectric metal, \cite{Ngai_main, Kiseliov_Feigelman, Volkov_ferro_super}.
At $T_{\mathrm{BG}},\omega_{\mathrm{TO}}> T$, estimating $\mathrm{coth}(\omega_{\mathbf{k}}/2T) \approx 1$, we obtain
\begin{align}\label{two_phonon_potential}\nonumber
V_2( \mathbf{q}) =
&- \left( \frac{g\Omega_0^2}{2\pi} \right)^2\int_{\mathbf{k}} \left\{1 + \frac{[\mathbf{k}\cdot(\mathbf{k}+ \mathbf{q})]^2}{\mathbf{k}^2 |\mathbf{k}+ \mathbf{q}|^2}\right\}
\\
&\times
\frac{1}{\omega_{\mathbf{k}}\omega_{\mathbf{k}+ \mathbf{q}}}\frac{1}{\omega_{\mathbf{k}}+ \omega_{\mathbf{k}+ \mathbf{q}}}.
\end{align}
The first factor under the integral originates from the transverse polarization of phonons, while the terms on the second line originate from the dispersion dependence of the phonon Green function.
To estimate the transition temperature, we consider (\ref{two_phonon_potential}) in the long-wave limit. Setting $V_2( 0)$, we reproduce previous result \cite{Kiseliov_Feigelman, Volkov_ferro_super}:
\begin{align}\label{Tc_2phonon}\nonumber
&
T_{c} \propto \mu e^{-1/\lambda_2\nu},\\
&
\lambda_2 = \left( \frac{g\Omega_0^2}{2\pi} \right)^2 \frac{1}{2s^3}
\ln\left[\frac{sq_0}{\max(T_{\mathrm{BG}},\omega_{\mathrm{TO}})}\right],
\end{align}
where $q_0$ is the large momentum cut-off which is determined by the lattice spacing.
 It was noted that as the paraelectric system is tuned closer to the ferroelectric instability, the softening of TO phonon gap might 
enhance the superconducting transition temperature, \cite{Kiseliov_Feigelman, Volkov_ferro_super}. We also note that although due to $\lambda_2\nu\propto T_{\mathrm{BG}} \ln[sq_0/T_{\mathrm{BG}}]$, the dome-like shape dependence of the 
superconducting transition temperature on the electron density is expected \cite{Volkov_ferro_super}, it is rather beyond the assumed approximations of the theory \cite{Kiseliov_Feigelman}. 

We argue that in the ferroelectric metal one-phonon coupling processes shall be taken into account as well. 
The respective interaction term is given by the static part of Eq. (\ref{int1}), 
$
V_1(\mathbf{q}) = -\frac{1}{\pi \omega^2_{{\bf q}}} (g \Omega_0 P_0)^2
\sin^2\phi_{\mathbf{q} \mathbf{P}_0}
$.

We seek for the transition temperature to s-wave superconducting state.
Substituting $V_1(\mathbf{q})$ into Eq. \ref{SQE} and integrating the resulting equation over the directions of momentum $\mathbf{q}$, in the long wave limit we obtain 
\begin{align}\label{TC_main_result}\nonumber
&T_{c} \propto \mu \exp\left\{-\frac{1}{\left(\lambda_1+\lambda_2\right)\nu}\right\},\\
&\lambda_1 = \frac{2}{3\pi } \left( \frac{g\Omega_0 P_0}{T_{\mathrm{BG}}} \right)^2  \ln\left(\frac{\omega_{\mathrm{TO}}^2 + T_{\mathrm{BG}}^2}{\omega_{\mathrm{TO}}^2}\right).
\end{align}
Here we assumed $\omega_\mathrm{TO}>\frac{s}{v_F}T$.
As the system is tuned deep into the ferroelectric state, the phonon gap $\omega_{\mathrm{TO}}$ increases and, hence, the two-phonon contribution to the interaction potential decreases logarithmically, in accordance with (\ref{Tc_2phonon}). 
We note that just the two-phonon mechanism gives roughly the same $T_{\mathrm{c}}$ in the paraelectric and ferroelectric versions of the metal (we keep in mind STO \cite{Ferroelectric_superconductor_exp, Ferroelectric_T_exp, Ferroelectric_superconductor_exp_2,Tomioka2022}). This is because $\omega_{\mathrm{TO}}$ in the ferroelectric phase of ferroelectric metal increases and can become of the same value as that of the paraelectric metal.
On the other hand, the one-phonon mechanism in the ferroelectric phase of the metal adds up to the two-phonon's. The one-phonon mechanism might even be dominant at the superconducting transition temperature at $\omega_{\mathrm{TO}}\gg T_{\mathrm{BG}}$ provided $P_0/\omega_{\mathrm{TO}} \gtrsim (\Omega_0/4\pi s^{3/2})\sqrt{\ln|sq_0/\omega_{\mathrm{TO}}|}$. This is consistent with the experiments \cite{Ferroelectric_superconductor_exp, Ferroelectric_T_exp, Ferroelectric_superconductor_exp_2,Tomioka2022} which observe enhancement of $T_{\mathrm{c}}$ inside the ferroelectric phase of the ferroelectric STO as compared to the paraelectric STO. 

It is instructive to comment on the density of states dependence of the one-phonon contribution to the transition temperature, noting 
$\lambda_1\nu \propto T_{\mathrm{BG}}^{-1}\ln(1+ T^2_{\mathrm{BG}}/\omega^2_{\mathrm{TO}})$. At $\omega_{\mathrm{TO}}> T_{\mathrm{BG}}$, the exponent in (\ref{TC_main_result}) follows the usual BCS dependence on the density of states. However, at $T_{\mathrm{BG}}>\omega_{\mathrm{TO}}$, with the logarithmic accuracy the interaction constant is inversely proportional to the Bloch-Gruneisen temperature squared, hence $\lambda_1\nu \propto \nu/T_{\mathrm{BG}}^2\propto m/k_F$. Surprisingly, in this case the decrease of the doping might enhance one-phonon contribution and increase $T_c$. 
Based on our findings, the superconducting transition temperature might have dome-like shape as a function of carrier concentration in the ferroelectric phase due to the one-phonon contribution. 

We note in passing that by approximating angular dependence of one-phonon coupling Eq. (\ref{int1}) by its average (see also \cite{SM_editors}), we have provided arguments for the isotropic superconductivity in the ferroelectric phase. In general, the superconductivity in ferroelectric phase will be anisotropic.
Moreover, ferroelectric phase will contain domains with different directions of the electric polarization, which might influence the superconducting temperature \cite{Hameed2022}. Both questions are left for future research.

%---------------------------------------------------------------------------------------------------------------------
\textit{Conclusions.} 
%---------------------------------------------------------------------------------------------------------------------
To conclude, we showed that a new mechanism of electron interaction with one TO-phonon emerges in the ferroelectric phase of the metal, compared to a well-known \cite{Ngai_main, ELL_conductivity_1, ELL_conductivity_2, Maslov_conductivity, Feigelman_conductivity, Marel_2phonon, Kiseliov_Feigelman, Volkov_ferro_super} two-phonon mechanism in the paraelectric phase.
We calculated the temperature dependence of resistivity and predicted anisotropic electric current response as one of the 
smoking gun signatures of the ferroelectric polarization onset.
We also analyzed the superconducting transition temperature in the ferroelectric phase. 
We think that our results for the temperature dependence of the resistivity in the vicinity of the ferroelectric transition given in Eqs. (\ref{result1}) and (\ref{result2}), as well as for the increase of the superconducting transition temperature given in Eq. (\ref{TC_main_result})
qualitatively agree with the experimentally observed ones in Sr$_{1-x}$Ca$_{x}$TiO$_{3}$ systems \cite{Ferroelectric_superconductor_exp, Ferroelectric_T_exp, Ferroelectric_superconductor_exp_2,Tomioka2022}. 
In particular, the one-phonon scattering mechanism explains the experimentally observed dip in the resistivity at the ferroelectric transition temperature, and the enhancement of the superconducting transition temperature in the ferroelectric STO.

%---------------------------------------------------------------------------------------------------------------------
\textit{Acknowledgments.} 
%---------------------------------------------------------------------------------------------------------------------
We would like to thank the Pirinem School of Theoretical Physics, where all work was initiated for the warm hospitality. 
We thank A.~T.~Burkov and A.~Yu.~Zyuzin for helpful discussions. 
VAZ thanks Abhishek Kumar and D.~L.~Maslov for hospitality during his stay at the UF. 
VAZ is supported by the Russian Foundation for Basic Research (grant No. 20-52-12
013), Deutsche Forschungsgemeinschaft (grant No. EV 30/14-1) cooperation, and by the Foundation for the Advancement of Theoretical Physics and Mathematics BASIS.
AAZ is supported by the Academy of Finland (project 308339) and in parts by the Academy of Finland Centre of Excellence program (project 336810).

%%%%%%%%%%%%%%%%%%%%%%%%%%%%%%%%%%%%%%%%%%%%%%%%%%%%%%%%%%%%%%%%%%%%%
%%%%%%%%%%%%%%%%%%%%%%%%%%%%%%%%%%%%%%%%%%%%%%%%%%%%%%%%%%%%%%%%%%%%%
%%%% Bibliography
%\bibliographystyle{apsrev}
\bibliography{Ferro_sc_references_arxiv}

\begin{thebibliography}{33}%
\makeatletter
\providecommand \@ifxundefined [1]{%
 \@ifx{#1\undefined}
}%
\providecommand \@ifnum [1]{%
 \ifnum #1\expandafter \@firstoftwo
 \else \expandafter \@secondoftwo
 \fi
}%
\providecommand \@ifx [1]{%
 \ifx #1\expandafter \@firstoftwo
 \else \expandafter \@secondoftwo
 \fi
}%
\providecommand \natexlab [1]{#1}%
\providecommand \enquote  [1]{``#1''}%
\providecommand \bibnamefont  [1]{#1}%
\providecommand \bibfnamefont [1]{#1}%
\providecommand \citenamefont [1]{#1}%
\providecommand \href@noop [0]{\@secondoftwo}%
\providecommand \href [0]{\begingroup \@sanitize@url \@href}%
\providecommand \@href[1]{\@@startlink{#1}\@@href}%
\providecommand \@@href[1]{\endgroup#1\@@endlink}%
\providecommand \@sanitize@url [0]{\catcode `\\12\catcode `\$12\catcode
  `\&12\catcode `\#12\catcode `\^12\catcode `\_12\catcode `\%12\relax}%
\providecommand \@@startlink[1]{}%
\providecommand \@@endlink[0]{}%
\providecommand \url  [0]{\begingroup\@sanitize@url \@url }%
\providecommand \@url [1]{\endgroup\@href {#1}{\urlprefix }}%
\providecommand \urlprefix  [0]{URL }%
\providecommand \Eprint [0]{\href }%
\providecommand \doibase [0]{https://doi.org/}%
\providecommand \selectlanguage [0]{\@gobble}%
\providecommand \bibinfo  [0]{\@secondoftwo}%
\providecommand \bibfield  [0]{\@secondoftwo}%
\providecommand \translation [1]{[#1]}%
\providecommand \BibitemOpen [0]{}%
\providecommand \bibitemStop [0]{}%
\providecommand \bibitemNoStop [0]{.\EOS\space}%
\providecommand \EOS [0]{\spacefactor3000\relax}%
\providecommand \BibitemShut  [1]{\csname bibitem#1\endcsname}%
\let\auto@bib@innerbib\@empty
%</preamble>
\bibitem [{\citenamefont {Gastiasoro}\ \emph {et~al.}(2020)\citenamefont
  {Gastiasoro}, \citenamefont {Ruhman},\ and\ \citenamefont
  {Fernandes}}]{Fernandes_review}%
  \BibitemOpen
  \bibfield  {author} {\bibinfo {author} {\bibfnamefont {M.~N.}\ \bibnamefont
  {Gastiasoro}}, \bibinfo {author} {\bibfnamefont {J.}~\bibnamefont {Ruhman}},\
  and\ \bibinfo {author} {\bibfnamefont {R.~M.}\ \bibnamefont {Fernandes}},\
  }\bibfield  {title} {\bibinfo {title} {{Superconductivity in dilute
  ${\mathrm{SrTiO}}_{3}$: A review}},\ }\href
  {https://doi.org/https://doi.org/10.1016/j.aop.2020.168107} {\bibfield
  {journal} {\bibinfo  {journal} {Annals of Physics}\ }\textbf {\bibinfo
  {volume} {417}},\ \bibinfo {pages} {168107} (\bibinfo {year} {2020})},\
  \bibinfo {note} {eliashberg theory at 60: Strong-coupling superconductivity
  and beyond}\BibitemShut {NoStop}%
\bibitem [{\citenamefont {Collignon}\ \emph {et~al.}(2019)\citenamefont
  {Collignon}, \citenamefont {Lin}, \citenamefont {Rischau}, \citenamefont
  {Fauque},\ and\ \citenamefont {Behnia}}]{Behnia_review}%
  \BibitemOpen
  \bibfield  {author} {\bibinfo {author} {\bibfnamefont {C.}~\bibnamefont
  {Collignon}}, \bibinfo {author} {\bibfnamefont {X.}~\bibnamefont {Lin}},
  \bibinfo {author} {\bibfnamefont {C.~W.}\ \bibnamefont {Rischau}}, \bibinfo
  {author} {\bibfnamefont {B.}~\bibnamefont {Fauque}},\ and\ \bibinfo {author}
  {\bibfnamefont {K.}~\bibnamefont {Behnia}},\ }\bibfield  {title} {\bibinfo
  {title} {{Metallicity and Superconductivity in Doped Strontium Titanate}},\
  }\href {https://doi.org/10.1146/annurev-conmatphys-031218-013144} {\bibfield
  {journal} {\bibinfo  {journal} {Annual Review of Condensed Matter Physics}\
  }\textbf {\bibinfo {volume} {10}},\ \bibinfo {pages} {25} (\bibinfo {year}
  {2019})}\BibitemShut {NoStop}%
\bibitem [{\citenamefont {Scheerer}\ \emph {et~al.}(2019)\citenamefont
  {Scheerer}, \citenamefont {Boselli}, \citenamefont {Pulmannova},
  \citenamefont {Rischau}, \citenamefont {Waelchli}, \citenamefont {Gariglio},
  \citenamefont {Giannini}, \citenamefont {van~der Marel},\ and\ \citenamefont
  {Triscone}}]{Muller_review}%
  \BibitemOpen
  \bibfield  {author} {\bibinfo {author} {\bibfnamefont {G.}~\bibnamefont
  {Scheerer}}, \bibinfo {author} {\bibfnamefont {M.}~\bibnamefont {Boselli}},
  \bibinfo {author} {\bibfnamefont {D.}~\bibnamefont {Pulmannova}}, \bibinfo
  {author} {\bibfnamefont {C.~W.}\ \bibnamefont {Rischau}}, \bibinfo {author}
  {\bibfnamefont {A.}~\bibnamefont {Waelchli}}, \bibinfo {author}
  {\bibfnamefont {S.}~\bibnamefont {Gariglio}}, \bibinfo {author}
  {\bibfnamefont {E.}~\bibnamefont {Giannini}}, \bibinfo {author}
  {\bibfnamefont {D.}~\bibnamefont {van~der Marel}},\ and\ \bibinfo {author}
  {\bibfnamefont {J.-M.}\ \bibnamefont {Triscone}},\ }\bibfield  {title}
  {\bibinfo {title} {{Ferroelectricity, Superconductivity, and
  ${\mathrm{SrTiO}}_{3}$ - Passions of K.A. Muller}},\ }\href
  {https://doi.org/doi:10.3390/condmat5040060} {\bibfield  {journal} {\bibinfo
  {journal} {Condens. Matter}\ }\textbf {\bibinfo {volume} {5}},\ \bibinfo
  {pages} {60} (\bibinfo {year} {2019})}\BibitemShut {NoStop}%
\bibitem [{\citenamefont {Bednorz}\ and\ \citenamefont
  {M\"uller}(1984)}]{Bednorz_Ca}%
  \BibitemOpen
  \bibfield  {author} {\bibinfo {author} {\bibfnamefont {J.~G.}\ \bibnamefont
  {Bednorz}}\ and\ \bibinfo {author} {\bibfnamefont {K.~A.}\ \bibnamefont
  {M\"uller}},\ }\bibfield  {title} {\bibinfo {title}
  {{${\mathrm{Sr}}_{1\ensuremath{-}x}{\mathrm{Ca}}_{x}\mathrm{Ti}{\mathrm{O}}_{3}$:
  An $\mathrm{XY}$ Quantum Ferroelectric with Transition to Randomness}},\
  }\href {https://doi.org/10.1103/PhysRevLett.52.2289} {\bibfield  {journal}
  {\bibinfo  {journal} {Phys. Rev. Lett.}\ }\textbf {\bibinfo {volume} {52}},\
  \bibinfo {pages} {2289} (\bibinfo {year} {1984})}\BibitemShut {NoStop}%
\bibitem [{\citenamefont {Lemanov}\ \emph {et~al.}(1996)\citenamefont
  {Lemanov}, \citenamefont {Smirnova}, \citenamefont {Syrnikov},\ and\
  \citenamefont {Tarakanov}}]{Lemanov_Ba}%
  \BibitemOpen
  \bibfield  {author} {\bibinfo {author} {\bibfnamefont {V.~V.}\ \bibnamefont
  {Lemanov}}, \bibinfo {author} {\bibfnamefont {E.~P.}\ \bibnamefont
  {Smirnova}}, \bibinfo {author} {\bibfnamefont {P.~P.}\ \bibnamefont
  {Syrnikov}},\ and\ \bibinfo {author} {\bibfnamefont {E.~A.}\ \bibnamefont
  {Tarakanov}},\ }\bibfield  {title} {\bibinfo {title} {{Phase transitions and
  glasslike behavior in
  ${\mathrm{Sr}}_{1\mathrm{\ensuremath{-}}\mathit{x}}$${\mathrm{Ba}}_{\mathit{x}}$${\mathrm{TiO}}_{3}$}},\
  }\href {https://doi.org/10.1103/PhysRevB.54.3151} {\bibfield  {journal}
  {\bibinfo  {journal} {Phys. Rev. B}\ }\textbf {\bibinfo {volume} {54}},\
  \bibinfo {pages} {3151} (\bibinfo {year} {1996})}\BibitemShut {NoStop}%
\bibitem [{\citenamefont {Lemanov}\ \emph {et~al.}(1997)\citenamefont
  {Lemanov}, \citenamefont {Smirnova},\ and\ \citenamefont
  {Tarakanov}}]{Lemanov_Pb}%
  \BibitemOpen
  \bibfield  {author} {\bibinfo {author} {\bibfnamefont {V.~V.}\ \bibnamefont
  {Lemanov}}, \bibinfo {author} {\bibfnamefont {E.~P.}\ \bibnamefont
  {Smirnova}},\ and\ \bibinfo {author} {\bibfnamefont {E.~A.}\ \bibnamefont
  {Tarakanov}},\ }\bibfield  {title} {\bibinfo {title} {{Ferroelectric
  properties of $\mathrm{SrTiO}_3$-$\mathrm{PbTiO}_3$ solid solutions}},\
  }\href {https://doi.org/10.1134/1.1129917} {\bibfield  {journal} {\bibinfo
  {journal} {Phys. Solid State}\ }\textbf {\bibinfo {volume} {39}},\ \bibinfo
  {pages} {628} (\bibinfo {year} {1997})}\BibitemShut {NoStop}%
\bibitem [{\citenamefont {Itoh}\ \emph {et~al.}(1999)\citenamefont {Itoh},
  \citenamefont {Wang}, \citenamefont {Inaguma}, \citenamefont {Yamaguchi},
  \citenamefont {Shan},\ and\ \citenamefont
  {Nakamura}}]{Itoh_para_ferro_isotop}%
  \BibitemOpen
  \bibfield  {author} {\bibinfo {author} {\bibfnamefont {M.}~\bibnamefont
  {Itoh}}, \bibinfo {author} {\bibfnamefont {R.}~\bibnamefont {Wang}}, \bibinfo
  {author} {\bibfnamefont {Y.}~\bibnamefont {Inaguma}}, \bibinfo {author}
  {\bibfnamefont {T.}~\bibnamefont {Yamaguchi}}, \bibinfo {author}
  {\bibfnamefont {Y.-J.}\ \bibnamefont {Shan}},\ and\ \bibinfo {author}
  {\bibfnamefont {T.}~\bibnamefont {Nakamura}},\ }\bibfield  {title} {\bibinfo
  {title} {{Ferroelectricity Induced by Oxygen Isotope Exchange in Strontium
  Titanate Perovskite}},\ }\href {https://doi.org/10.1103/PhysRevLett.82.3540}
  {\bibfield  {journal} {\bibinfo  {journal} {Phys. Rev. Lett.}\ }\textbf
  {\bibinfo {volume} {82}},\ \bibinfo {pages} {3540} (\bibinfo {year}
  {1999})}\BibitemShut {NoStop}%
\bibitem [{\citenamefont {Uwe}\ and\ \citenamefont
  {Sakudo}(1976)}]{Stress_para_ferr_trans}%
  \BibitemOpen
  \bibfield  {author} {\bibinfo {author} {\bibfnamefont {H.}~\bibnamefont
  {Uwe}}\ and\ \bibinfo {author} {\bibfnamefont {T.}~\bibnamefont {Sakudo}},\
  }\bibfield  {title} {\bibinfo {title} {{Stress-induced ferroelectricity and
  soft phonon modes in SrTi${\mathrm{O}}_{3}$}},\ }\href
  {https://doi.org/10.1103/PhysRevB.13.271} {\bibfield  {journal} {\bibinfo
  {journal} {Phys. Rev. B}\ }\textbf {\bibinfo {volume} {13}},\ \bibinfo
  {pages} {271} (\bibinfo {year} {1976})}\BibitemShut {NoStop}%
\bibitem [{\citenamefont {Behnia}()}]{Behnia_T2_review}%
  \BibitemOpen
  \bibfield  {author} {\bibinfo {author} {\bibfnamefont {K.}~\bibnamefont
  {Behnia}},\ }\bibfield  {title} {\bibinfo {title} {{On the origin and the
  amplitude of T-square resistivity in Fermi liquids}},\ }\href
  {https://arxiv.org/abs/2112.11092} {\bibinfo  {journal} {arxiv: 2112.11092}\
  }\BibitemShut {NoStop}%
\bibitem [{\citenamefont {Schooley}\ \emph {et~al.}(1964)\citenamefont
  {Schooley}, \citenamefont {Hosler},\ and\ \citenamefont
  {Cohen}}]{Schooley_first_experiment}%
  \BibitemOpen
\bibfield  {journal} {  }\bibfield  {author} {\bibinfo {author} {\bibfnamefont
  {J.~F.}\ \bibnamefont {Schooley}}, \bibinfo {author} {\bibfnamefont {W.~R.}\
  \bibnamefont {Hosler}},\ and\ \bibinfo {author} {\bibfnamefont {M.~L.}\
  \bibnamefont {Cohen}},\ }\bibfield  {title} {\bibinfo {title}
  {{Superconductivity in Semiconducting SrTi${\mathrm{O}}_{3}$}},\ }\href
  {https://doi.org/10.1103/PhysRevLett.12.474} {\bibfield  {journal} {\bibinfo
  {journal} {Phys. Rev. Lett.}\ }\textbf {\bibinfo {volume} {12}},\ \bibinfo
  {pages} {474} (\bibinfo {year} {1964})}\BibitemShut {NoStop}%
\bibitem [{\citenamefont {Ruhman}\ and\ \citenamefont
  {Lee}(2016)}]{Ruhman_Lee}%
  \BibitemOpen
  \bibfield  {author} {\bibinfo {author} {\bibfnamefont {J.}~\bibnamefont
  {Ruhman}}\ and\ \bibinfo {author} {\bibfnamefont {P.~A.}\ \bibnamefont
  {Lee}},\ }\bibfield  {title} {\bibinfo {title} {{Superconductivity at very
  low density: The case of strontium titanate}},\ }\href
  {https://doi.org/10.1103/PhysRevB.94.224515} {\bibfield  {journal} {\bibinfo
  {journal} {Phys. Rev. B}\ }\textbf {\bibinfo {volume} {94}},\ \bibinfo
  {pages} {224515} (\bibinfo {year} {2016})}\BibitemShut {NoStop}%
\bibitem [{\citenamefont {Ngai}(1974)}]{Ngai_main}%
  \BibitemOpen
  \bibfield  {author} {\bibinfo {author} {\bibfnamefont {K.~L.}\ \bibnamefont
  {Ngai}},\ }\bibfield  {title} {\bibinfo {title} {{Two-Phonon Deformation
  Potential and Superconductivity in Degenerate Semiconductors}},\ }\href
  {https://doi.org/10.1103/PhysRevLett.32.215} {\bibfield  {journal} {\bibinfo
  {journal} {Phys. Rev. Lett.}\ }\textbf {\bibinfo {volume} {32}},\ \bibinfo
  {pages} {215} (\bibinfo {year} {1974})}\BibitemShut {NoStop}%
\bibitem [{\citenamefont {Epifanov}\ \emph {et~al.}(1981)\citenamefont
  {Epifanov}, \citenamefont {Levanyuk},\ and\ \citenamefont
  {Levanyuk}}]{ELL_conductivity_1}%
  \BibitemOpen
  \bibfield  {author} {\bibinfo {author} {\bibfnamefont {Y.~N.}\ \bibnamefont
  {Epifanov}}, \bibinfo {author} {\bibfnamefont {A.~P.}\ \bibnamefont
  {Levanyuk}},\ and\ \bibinfo {author} {\bibfnamefont {G.~M.}\ \bibnamefont
  {Levanyuk}},\ }\bibfield  {title} {\bibinfo {title} {{Interaction of carriers
  with TO phonons and electrical conductivity of ferroelectrics}},\ }\href
  {https://doi.org/10.1080/00150198108017687} {\bibfield  {journal} {\bibinfo
  {journal} {Ferroelectrics}\ }\textbf {\bibinfo {volume} {35}},\ \bibinfo
  {pages} {199} (\bibinfo {year} {1981})}\BibitemShut {NoStop}%
\bibitem [{\citenamefont {Epifanov}\ \emph {et~al.}(1982)\citenamefont
  {Epifanov}, \citenamefont {Levanyuk},\ and\ \citenamefont
  {Levanyuk}}]{ELL_conductivity_2}%
  \BibitemOpen
  \bibfield  {author} {\bibinfo {author} {\bibfnamefont {Y.~N.}\ \bibnamefont
  {Epifanov}}, \bibinfo {author} {\bibfnamefont {A.~P.}\ \bibnamefont
  {Levanyuk}},\ and\ \bibinfo {author} {\bibfnamefont {G.~M.}\ \bibnamefont
  {Levanyuk}},\ }\bibfield  {title} {\bibinfo {title} {{Interaction of carriers
  with soft ferroelectric mode and temperature dependence of conductivity}},\
  }\href {https://doi.org/10.1080/00150198208210644} {\bibfield  {journal}
  {\bibinfo  {journal} {Ferroelectrics}\ }\textbf {\bibinfo {volume} {43}},\
  \bibinfo {pages} {191} (\bibinfo {year} {1982})}\BibitemShut {NoStop}%
\bibitem [{\citenamefont {Stucky}\ \emph {et~al.}(2016)\citenamefont {Stucky},
  \citenamefont {Scheerer}, \citenamefont {Ren}, \citenamefont {Jaccard},
  \citenamefont {Poumirol}, \citenamefont {Barreteau}, \citenamefont
  {Giannini},\ and\ \citenamefont {van~der
  Marel}}]{Paraelectric_superconductor_exp}%
  \BibitemOpen
  \bibfield  {author} {\bibinfo {author} {\bibfnamefont {A.}~\bibnamefont
  {Stucky}}, \bibinfo {author} {\bibfnamefont {G.~W.}\ \bibnamefont
  {Scheerer}}, \bibinfo {author} {\bibfnamefont {Z.}~\bibnamefont {Ren}},
  \bibinfo {author} {\bibfnamefont {D.}~\bibnamefont {Jaccard}}, \bibinfo
  {author} {\bibfnamefont {J.~M.}\ \bibnamefont {Poumirol}}, \bibinfo {author}
  {\bibfnamefont {C.}~\bibnamefont {Barreteau}}, \bibinfo {author}
  {\bibfnamefont {E.}~\bibnamefont {Giannini}},\ and\ \bibinfo {author}
  {\bibfnamefont {D.}~\bibnamefont {van~der Marel}},\ }\bibfield  {title}
  {\bibinfo {title} {{Isotope effect in superconducting n-doped
  ${\mathrm{SrTiO}}_{3}$}},\ }\href {https://doi.org/10.1038/srep37582}
  {\bibfield  {journal} {\bibinfo  {journal} {Scientific Reports}\ }\textbf
  {\bibinfo {volume} {6}},\ \bibinfo {pages} {37582} (\bibinfo {year}
  {2016})}\BibitemShut {NoStop}%
\bibitem [{\citenamefont {Kumar}\ \emph {et~al.}(2021)\citenamefont {Kumar},
  \citenamefont {Yudson},\ and\ \citenamefont {Maslov}}]{Maslov_conductivity}%
  \BibitemOpen
  \bibfield  {author} {\bibinfo {author} {\bibfnamefont {A.}~\bibnamefont
  {Kumar}}, \bibinfo {author} {\bibfnamefont {V.~I.}\ \bibnamefont {Yudson}},\
  and\ \bibinfo {author} {\bibfnamefont {D.~L.}\ \bibnamefont {Maslov}},\
  }\bibfield  {title} {\bibinfo {title} {{Quasiparticle and Nonquasiparticle
  Transport in Doped Quantum Paraelectrics}},\ }\href
  {https://doi.org/10.1103/PhysRevLett.126.076601} {\bibfield  {journal}
  {\bibinfo  {journal} {Phys. Rev. Lett.}\ }\textbf {\bibinfo {volume} {126}},\
  \bibinfo {pages} {076601} (\bibinfo {year} {2021})}\BibitemShut {NoStop}%
\bibitem [{\citenamefont {Nazaryan}\ and\ \citenamefont
  {Feigel'man}(2021)}]{Feigelman_conductivity}%
  \BibitemOpen
  \bibfield  {author} {\bibinfo {author} {\bibfnamefont {K.~G.}\ \bibnamefont
  {Nazaryan}}\ and\ \bibinfo {author} {\bibfnamefont {M.~V.}\ \bibnamefont
  {Feigel'man}},\ }\bibfield  {title} {\bibinfo {title} {{Conductivity and
  thermoelectric coefficients of doped ${\mathrm{SrTiO}}_{3}$ at high
  temperatures}},\ }\href {https://doi.org/10.1103/PhysRevB.104.115201}
  {\bibfield  {journal} {\bibinfo  {journal} {Phys. Rev. B}\ }\textbf {\bibinfo
  {volume} {104}},\ \bibinfo {pages} {115201} (\bibinfo {year}
  {2021})}\BibitemShut {NoStop}%
\bibitem [{\citenamefont {van~der Marel}\ \emph {et~al.}(2019)\citenamefont
  {van~der Marel}, \citenamefont {Barantani},\ and\ \citenamefont
  {Rischau}}]{Marel_2phonon}%
  \BibitemOpen
  \bibfield  {author} {\bibinfo {author} {\bibfnamefont {D.}~\bibnamefont
  {van~der Marel}}, \bibinfo {author} {\bibfnamefont {F.}~\bibnamefont
  {Barantani}},\ and\ \bibinfo {author} {\bibfnamefont {C.~W.}\ \bibnamefont
  {Rischau}},\ }\bibfield  {title} {\bibinfo {title} {{Possible mechanism for
  superconductivity in doped ${\mathrm{SrTiO}}_{3}$}},\ }\href
  {https://doi.org/10.1103/PhysRevResearch.1.013003} {\bibfield  {journal}
  {\bibinfo  {journal} {Phys. Rev. Research}\ }\textbf {\bibinfo {volume}
  {1}},\ \bibinfo {pages} {013003} (\bibinfo {year} {2019})}\BibitemShut
  {NoStop}%
\bibitem [{\citenamefont {Kiselov}\ and\ \citenamefont
  {Feigel'man}(2021)}]{Kiseliov_Feigelman}%
  \BibitemOpen
  \bibfield  {author} {\bibinfo {author} {\bibfnamefont {D.~E.}\ \bibnamefont
  {Kiselov}}\ and\ \bibinfo {author} {\bibfnamefont {M.~V.}\ \bibnamefont
  {Feigel'man}},\ }\bibfield  {title} {\bibinfo {title} {{Theory of
  superconductivity due to Ngai's mechanism in lightly doped
  ${\mathrm{SrTiO}}_{3}$}},\ }\href
  {https://doi.org/10.1103/PhysRevB.104.L220506} {\bibfield  {journal}
  {\bibinfo  {journal} {Phys. Rev. B}\ }\textbf {\bibinfo {volume} {104}},\
  \bibinfo {pages} {L220506} (\bibinfo {year} {2021})}\BibitemShut {NoStop}%
\bibitem [{\citenamefont {Volkov}\ \emph {et~al.}(2022)\citenamefont {Volkov},
  \citenamefont {Chandra},\ and\ \citenamefont {Coleman}}]{Volkov_ferro_super}%
  \BibitemOpen
  \bibfield  {author} {\bibinfo {author} {\bibfnamefont {P.~A.}\ \bibnamefont
  {Volkov}}, \bibinfo {author} {\bibfnamefont {P.}~\bibnamefont {Chandra}},\
  and\ \bibinfo {author} {\bibfnamefont {P.}~\bibnamefont {Coleman}},\
  }\bibfield  {title} {\bibinfo {title} {{Superconductivity from energy
  fluctuations in dilute quantum critical polar metals}},\ }\href
  {https://www.nature.com/articles/s41467-022-32303-2} {\bibfield  {journal}
  {\bibinfo  {journal} {Nature Communications}\ }\textbf {\bibinfo {volume}
  {13}},\ \bibinfo {pages} {4599} (\bibinfo {year} {2022})}\BibitemShut
  {NoStop}%
\bibitem [{\citenamefont {Rischau}\ \emph {et~al.}(2017)\citenamefont
  {Rischau}, \citenamefont {Lin}, \citenamefont {Grams}, \citenamefont {Finck},
  \citenamefont {Harms}, \citenamefont {Engelmayer}, \citenamefont {Lorenz},
  \citenamefont {Gallais}, \citenamefont {Fauquy}, \citenamefont {Hemberger},\
  and\ \citenamefont {Behnia}}]{Ferroelectric_superconductor_exp}%
  \BibitemOpen
  \bibfield  {author} {\bibinfo {author} {\bibfnamefont {C.~W.}\ \bibnamefont
  {Rischau}}, \bibinfo {author} {\bibfnamefont {X.}~\bibnamefont {Lin}},
  \bibinfo {author} {\bibfnamefont {C.~P.}\ \bibnamefont {Grams}}, \bibinfo
  {author} {\bibfnamefont {D.}~\bibnamefont {Finck}}, \bibinfo {author}
  {\bibfnamefont {S.}~\bibnamefont {Harms}}, \bibinfo {author} {\bibfnamefont
  {J.}~\bibnamefont {Engelmayer}}, \bibinfo {author} {\bibfnamefont
  {T.}~\bibnamefont {Lorenz}}, \bibinfo {author} {\bibfnamefont
  {Y.}~\bibnamefont {Gallais}}, \bibinfo {author} {\bibfnamefont
  {B.}~\bibnamefont {Fauquy}}, \bibinfo {author} {\bibfnamefont
  {J.}~\bibnamefont {Hemberger}},\ and\ \bibinfo {author} {\bibfnamefont
  {K.}~\bibnamefont {Behnia}},\ }\bibfield  {title} {\bibinfo {title} {{A
  ferroelectric quantum phase transition inside the superconducting dome of
  $\mathrm{Sr}_{1-x}\mathrm{Ca}_x\mathrm{TiO}_{3-\delta}$}},\ }\href
  {https://doi.org/10.1038/nphys4085} {\bibfield  {journal} {\bibinfo
  {journal} {Nature Phys.}\ }\textbf {\bibinfo {volume} {13}},\ \bibinfo
  {pages} {643} (\bibinfo {year} {2017})}\BibitemShut {NoStop}%
\bibitem [{\citenamefont {Wang}\ \emph {et~al.}(2019)\citenamefont {Wang},
  \citenamefont {Yang}, \citenamefont {Rischau}, \citenamefont {Xu},
  \citenamefont {Ren}, \citenamefont {Lorenz}, \citenamefont {Hemberger},
  \citenamefont {Lin},\ and\ \citenamefont {Behnia}}]{Ferroelectric_T_exp}%
  \BibitemOpen
  \bibfield  {author} {\bibinfo {author} {\bibfnamefont {J.}~\bibnamefont
  {Wang}}, \bibinfo {author} {\bibfnamefont {L.}~\bibnamefont {Yang}}, \bibinfo
  {author} {\bibfnamefont {C.~W.}\ \bibnamefont {Rischau}}, \bibinfo {author}
  {\bibfnamefont {Z.}~\bibnamefont {Xu}}, \bibinfo {author} {\bibfnamefont
  {Z.}~\bibnamefont {Ren}}, \bibinfo {author} {\bibfnamefont {T.}~\bibnamefont
  {Lorenz}}, \bibinfo {author} {\bibfnamefont {J.}~\bibnamefont {Hemberger}},
  \bibinfo {author} {\bibfnamefont {X.}~\bibnamefont {Lin}},\ and\ \bibinfo
  {author} {\bibfnamefont {K.}~\bibnamefont {Behnia}},\ }\bibfield  {title}
  {\bibinfo {title} {{Charge transport in a polar metal}},\ }\href
  {https://doi.org/10.1038/s41535-019-0200-1} {\bibfield  {journal} {\bibinfo
  {journal} {npj Quantum Materials}\ }\textbf {\bibinfo {volume} {104}},\
  \bibinfo {pages} {61} (\bibinfo {year} {2019})}\BibitemShut {NoStop}%
\bibitem [{\citenamefont {Rischau}\ \emph {et~al.}(2022)\citenamefont
  {Rischau}, \citenamefont {Pulmannov\'a}, \citenamefont {Scheerer},
  \citenamefont {Stucky}, \citenamefont {Giannini},\ and\ \citenamefont
  {van~der Marel}}]{Ferroelectric_superconductor_exp_2}%
  \BibitemOpen
  \bibfield  {author} {\bibinfo {author} {\bibfnamefont {C.~W.}\ \bibnamefont
  {Rischau}}, \bibinfo {author} {\bibfnamefont {D.}~\bibnamefont
  {Pulmannov\'a}}, \bibinfo {author} {\bibfnamefont {G.~W.}\ \bibnamefont
  {Scheerer}}, \bibinfo {author} {\bibfnamefont {A.}~\bibnamefont {Stucky}},
  \bibinfo {author} {\bibfnamefont {E.}~\bibnamefont {Giannini}},\ and\
  \bibinfo {author} {\bibfnamefont {D.}~\bibnamefont {van~der Marel}},\
  }\bibfield  {title} {\bibinfo {title} {{Isotope tuning of the superconducting
  dome of strontium titanate}},\ }\href
  {https://doi.org/10.1103/PhysRevResearch.4.013019} {\bibfield  {journal}
  {\bibinfo  {journal} {Phys. Rev. Research}\ }\textbf {\bibinfo {volume}
  {4}},\ \bibinfo {pages} {013019} (\bibinfo {year} {2022})}\BibitemShut
  {NoStop}%
\bibitem [{\citenamefont {Tomioka}\ \emph {et~al.}()\citenamefont {Tomioka},
  \citenamefont {Shirakawa},\ and\ \citenamefont {Inoue}}]{Tomioka2022}%
  \BibitemOpen
  \bibfield  {author} {\bibinfo {author} {\bibfnamefont {Y.}~\bibnamefont
  {Tomioka}}, \bibinfo {author} {\bibfnamefont {N.}~\bibnamefont {Shirakawa}},\
  and\ \bibinfo {author} {\bibfnamefont {I.~H.}\ \bibnamefont {Inoue}},\
  }\bibfield  {title} {\bibinfo {title} {{Superconductivity enhanced in the
  polar metal region of Sr$_{0.95}$Ba$_{0.05}$TiO$_3$ and
  Sr$_{0.985}$Ca$_{0.015}$TiO$_3$ revealed by the systematic Nb doping}},\
  }\href {https://arxiv.org/abs/2203.16208} {\bibinfo  {journal} {arxiv:
  2203.16208}\ }\BibitemShut {NoStop}%
\bibitem [{\citenamefont {Cohen}(1992)}]{Cohen_review}%
  \BibitemOpen
\bibfield  {journal} {  }\bibfield  {author} {\bibinfo {author} {\bibfnamefont
  {R.}~\bibnamefont {Cohen}},\ }\bibfield  {title} {\bibinfo {title} {{Origin
  of ferroelectricity in perovskite oxides}},\ }\href
  {https://doi.org/10.1038/358136a0} {\bibfield  {journal} {\bibinfo  {journal}
  {Nature}\ }\textbf {\bibinfo {volume} {358}},\ \bibinfo {pages} {136}
  (\bibinfo {year} {1992})}\BibitemShut {NoStop}%
\bibitem [{\citenamefont {Zhou}\ and\ \citenamefont
  {Ariando}(2020)}]{Ferroelectric_review}%
  \BibitemOpen
  \bibfield  {author} {\bibinfo {author} {\bibfnamefont {W.~X.}\ \bibnamefont
  {Zhou}}\ and\ \bibinfo {author} {\bibfnamefont {A.}~\bibnamefont {Ariando}},\
  }\bibfield  {title} {\bibinfo {title} {{Review on ferroelectric/polar
  metals}},\ }\href {https://doi.org/10.35848/1347-4065/ab8bbf} {\bibfield
  {journal} {\bibinfo  {journal} {Japanese Journal of Applied Physics}\
  }\textbf {\bibinfo {volume} {59}},\ \bibinfo {pages} {SI0802} (\bibinfo
  {year} {2020})}\BibitemShut {NoStop}%
\bibitem [{\citenamefont {Benedek}\ and\ \citenamefont
  {Birol}(2016)}]{Benedek}%
  \BibitemOpen
  \bibfield  {author} {\bibinfo {author} {\bibfnamefont {N.~A.}\ \bibnamefont
  {Benedek}}\ and\ \bibinfo {author} {\bibfnamefont {T.}~\bibnamefont
  {Birol}},\ }\bibfield  {title} {\bibinfo {title} {{`Ferroelectric' metals
  reexamined: fundamental mechanisms and design considerations for new
  materials}},\ }\href {https://doi.org/10.1039/C5TC03856A} {\bibfield
  {journal} {\bibinfo  {journal} {Journal of Material Chemistry C}\ }\textbf
  {\bibinfo {volume} {4}},\ \bibinfo {pages} {4000} (\bibinfo {year}
  {2016})}\BibitemShut {NoStop}%
\bibitem [{SM_()}]{SM_editors}%
  \BibitemOpen
  \href@noop {} {}\bibinfo {note} {{See Supplemental Material at [URL will be
  inserted by publisher]}}\BibitemShut {NoStop}%
\bibitem [{\citenamefont {Allen}\ and\ \citenamefont
  {Silberglitt}(1974)}]{Allan}%
  \BibitemOpen
  \bibfield  {author} {\bibinfo {author} {\bibfnamefont {P.~B.}\ \bibnamefont
  {Allen}}\ and\ \bibinfo {author} {\bibfnamefont {R.}~\bibnamefont
  {Silberglitt}},\ }\bibfield  {title} {\bibinfo {title} {{Some effects of
  phonon dynamics on electron lifetime, mass renormalization, and
  superconducting transition temperature}},\ }\href
  {https://doi.org/10.1103/PhysRevB.9.4733} {\bibfield  {journal} {\bibinfo
  {journal} {Phys. Rev. B}\ }\textbf {\bibinfo {volume} {9}},\ \bibinfo {pages}
  {4733} (\bibinfo {year} {1974})}\BibitemShut {NoStop}%
\bibitem [{\citenamefont {Migdal}(1958)}]{Migdal}%
  \BibitemOpen
  \bibfield  {author} {\bibinfo {author} {\bibfnamefont {A.~B.}\ \bibnamefont
  {Migdal}},\ }\bibfield  {title} {\bibinfo {title} {{Interaction between
  electrons and lattice vibrations in a normal metal}},\ }\href
  {http://jetp.ras.ru/cgi-bin/dn/e_007_06_0996.pdf} {\bibfield  {journal}
  {\bibinfo  {journal} {Sov. Phys. -JETP}\ }\textbf {\bibinfo {volume} {7}},\
  \bibinfo {pages} {996} (\bibinfo {year} {1958})}\BibitemShut {NoStop}%
\bibitem [{\citenamefont {Gor'kov}\ and\ \citenamefont
  {Melik-Barkhudarov}(1961)}]{Barkhudarov_Tc}%
  \BibitemOpen
  \bibfield  {author} {\bibinfo {author} {\bibfnamefont {L.~P.}\ \bibnamefont
  {Gor'kov}}\ and\ \bibinfo {author} {\bibfnamefont {T.~K.}\ \bibnamefont
  {Melik-Barkhudarov}},\ }\bibfield  {title} {\bibinfo {title} {{Contribution
  to the theory of superfluidity in an imperfect Fermi gas}},\ }\href
  {http://jetp.ras.ru/cgi-bin/dn/e_013_05_1018.pdf} {\bibfield  {journal}
  {\bibinfo  {journal} {Sov.Phys. - JETP}\ }\textbf {\bibinfo {volume} {40}},\
  \bibinfo {pages} {1452} (\bibinfo {year} {1961})}\BibitemShut {NoStop}%
\bibitem [{\citenamefont {Gor'kov}(2016)}]{Gorkov_Tc}%
  \BibitemOpen
  \bibfield  {author} {\bibinfo {author} {\bibfnamefont {L.~P.}\ \bibnamefont
  {Gor'kov}},\ }\bibfield  {title} {\bibinfo {title} {{Superconducting
  transition temperature: Interacting Fermi gas and phonon mechanisms in the
  nonadiabatic regime}},\ }\href {https://doi.org/10.1103/PhysRevB.93.054517}
  {\bibfield  {journal} {\bibinfo  {journal} {Phys. Rev. B}\ }\textbf {\bibinfo
  {volume} {93}},\ \bibinfo {pages} {054517} (\bibinfo {year}
  {2016})}\BibitemShut {NoStop}%
\bibitem [{\citenamefont {Hameed}\ \emph {et~al.}(2022)\citenamefont {Hameed},
  \citenamefont {Pelc}, \citenamefont {Anderson}, \citenamefont {Klein},
  \citenamefont {Spieker}, \citenamefont {Yue}, \citenamefont {Ramberger},
  \citenamefont {Lukas}, \citenamefont {Liu}, \citenamefont {Krogstad},
  \citenamefont {Osborn}, \citenamefont {Li}, \citenamefont {Leighton},
  \citenamefont {Fernandes},\ and\ \citenamefont {Greven}}]{Hameed2022}%
  \BibitemOpen
  \bibfield  {author} {\bibinfo {author} {\bibfnamefont {S.}~\bibnamefont
  {Hameed}}, \bibinfo {author} {\bibfnamefont {D.}~\bibnamefont {Pelc}},
  \bibinfo {author} {\bibfnamefont {Z.}~\bibnamefont {Anderson}}, \bibinfo
  {author} {\bibfnamefont {A.}~\bibnamefont {Klein}}, \bibinfo {author}
  {\bibfnamefont {R.~J.}\ \bibnamefont {Spieker}}, \bibinfo {author}
  {\bibfnamefont {B.}~\bibnamefont {Yue}, \bibfnamefont {L.~ans~Das}}, \bibinfo
  {author} {\bibfnamefont {J.}~\bibnamefont {Ramberger}}, \bibinfo {author}
  {\bibfnamefont {M.}~\bibnamefont {Lukas}}, \bibinfo {author} {\bibfnamefont
  {Y.}~\bibnamefont {Liu}}, \bibinfo {author} {\bibfnamefont {M.~J.}\
  \bibnamefont {Krogstad}}, \bibinfo {author} {\bibfnamefont {R.}~\bibnamefont
  {Osborn}}, \bibinfo {author} {\bibfnamefont {Y.}~\bibnamefont {Li}}, \bibinfo
  {author} {\bibfnamefont {C.}~\bibnamefont {Leighton}}, \bibinfo {author}
  {\bibfnamefont {R.~M.}\ \bibnamefont {Fernandes}},\ and\ \bibinfo {author}
  {\bibfnamefont {M.}~\bibnamefont {Greven}},\ }\bibfield  {title} {\bibinfo
  {title} {{Enhanced superconductivity and ferroelectric quantum criticality in
  plastically deformed strontium titanate}},\ }\href
  {https://doi.org/10.1038/s41563-021-01102-3} {\bibfield  {journal} {\bibinfo
  {journal} {Nature Materials}\ }\textbf {\bibinfo {volume} {21}},\ \bibinfo
  {pages} {54} (\bibinfo {year} {2022})}\BibitemShut {NoStop}%
\end{thebibliography}%

%%%%%%%%%%%%%%%%%%%%%%%%%%%%%%%%%%%%%%%%%%%%%%%%%%%%%%%%%%%%%%%%%%%%%
%%%% Supplementary material 

\clearpage
\onecolumngrid
\begin{center}
\rule{0.38\linewidth}{1pt}\\
\vspace{-0.37cm}\rule{0.49\linewidth}{1pt}
\end{center}
\setcounter{section}{0}
\setcounter{equation}{0}

\section*{Supplemental Material to \texorpdfstring{\\}{}
"Anisotropic resistivity and superconducting instability in ferroelectric metals"}

\begin{widetext}
%--------------------------------------------------------------------------------------------------------------------------------------------------------------
\subsection{Model of a ferroelectric metal}
%--------------------------------------------------------------------------------------------------------------------------------------------------------------

The Hamiltonian includes electrons, phonons and interaction between the two.
\begin{align}
H_{\mathrm{F}} = \int_{\bf r} \psi^{\dag}(x) \left( -\frac{\partial^2_{\bf r}}{2m } - \mu \right)\psi(x)
\end{align}
where $x=({\bf r},t)$ is a general coordinate, $\mu$ is the chemical potential, and $m$ is the mass of electrons.
Interaction of electrons with two phonons is
\begin{align}\label{ephSM}
H_{\mathrm{e}-2\mathrm{ph}} = g\int_{{\bf r}} \psi^{\dag}(x)\psi(x) \left[{\bf P}_{0}+{\bf P}(x)\right]\cdot\left[{\bf P}_{0}+{\bf P}(x)\right],
\end{align}
where $g$ is a constant.
Here ${\bf P}(x) $ is the phonon displacement field,
\begin{align}
{\bf P}(x) = \sum_{a=1,2}\sum_{{\bf q}} \frac{{\bf e}^{a}_{{\bf q}}}{\sqrt{V}}A_{\bf q}
\left[ b_{a;{\bf q}}(t) e^{i{\bf q}{\bf r}} + b_{a;{\bf q}}^{\dag}(t) e^{-i{\bf q}{\bf r}} \right],
\end{align}
where $A_{\bf q}^2 = \left[\epsilon_{0}({\bf q}) - \epsilon_{\infty} \right] \frac{ \omega_{\bf q}}{4\pi}$ with $\epsilon_{0}({\bf q}) = \frac{\Omega_{0}^2}{ \omega_{\bf q}^2}$ and $\epsilon_{0}({\bf q}) \gg \epsilon_{\infty}$, and $b_{a;{\bf q}}$ and $b_{a;{\bf q}}^{\dag}$ are boson fields. 
Polarization vectors ${\bf e}^{a}_{{\bf q}}$ of the two branches $a=1,2$ of transverse optical phonons satisfy
\begin{align}
\sum_{a=1,2} e^{a}_{\alpha;{\bf p}}e^{a}_{\beta;{\bf p}} = \delta_{\alpha\beta} - \frac{p_{\alpha}p_{\beta}}{p^2},
\end{align}
where $\alpha, ~\beta = x,y,z$. 
Hamiltonian of phonons is
\begin{align}
H^{(a)}_{\mathrm{ph}} = \int_{\bf q} \omega_{\bf q}\left(b_{a;{\bf q}}^{\dag}b_{a;{\bf q}} + \frac{1}{2}\right),
\end{align}
where $\omega_{\bf q} = \sqrt{\omega_{\mathrm{TO}}^2 + (sq)^2}$ is the dispersion of phonons with $\omega_{\mathrm{TO}}$ being the mass of phonons and $s$ being the speed of sound. 
Here $\int_{\bf q}(...) \equiv \int \frac{d{\bf q}}{(2\pi)^3} (...) $. Constant ${\bf P}_{0}$ in Eq. (\ref{ephSM}) is electric polarization occuring in a ferroelectric, and which can understood as a condensation of soft optical phonons when $\omega_{\mathrm{TO}} \rightarrow 0$ as the temperature is reduced to ferroelectric transition temperature, $T\rightarrow T_{\mathrm{C}}^{\mathrm{FE}}$. 
The condensation not only results in non-zero ${\bf P}_{0}$ but also in a finite gap $\omega_{\mathrm{TO}}$ below ferroelectric transition temperature, i.e. $T<T_{\mathrm{C}}^{\mathrm{FE}}$. The gap there is propotional to the absolute value of ${\bf P}_{0}$, i.e. $\omega_{\mathrm{TO}}\propto \vert {\bf P}_{0} \vert$.
We assume that the situation is such that the density of conducting electrons is low and they don't screen the finite polarization ${\bf P}_{0}$.

Although, the Matsubara technique is certainly the most conventional and historic choice to study superconducting instability of the system, here we wish to work in the Keldysh technique. The Keldysh technique has a power to describe not only equilibrium but also non-equilibrium systems, allows for a convenient derivation of the kinetic equation, and is probably the best in describing disordered electron systems (by avoiding replicas).
Therefore, we will make an effort to exercise the Keldysh technique. We will follow the book [1].

As far as notations are concerned, electron fields are promoted to Grassmann fields, $\psi^{\dag}(x) \rightarrow \bar{\psi}(x)$ and $\psi(x) \rightarrow \psi(x)$, in accord with construction of electronic path integral.
General Hamiltonian bilinear in electron operators reads as
\begin{align}
\int_{{\cal C}} dt~\bar{\psi}(t)\theta(t)\psi(t) 
= 
\int_{-\infty}^{+\infty}dt~ \hat{\bar{\psi}}(t)
\left[\begin{array}{cc} \theta_{+}(t) & 0 \\ 0 & - \theta_{-}(t)\end{array}\right]\hat{\psi}(t)
=
\int_{-\infty}^{+\infty}dt~ \hat{\bar{\Psi}}(t)\hat{L}
\left[\begin{array}{cc} \theta_{+}(t) & 0 \\ 0 & \theta_{-}(t)\end{array}\right]\hat{L}^{-1}\hat{\Psi}(t)
\end{align}
where 
\begin{align}
\hat{\bar{\psi}}=[ \bar{\psi}_{+}, ~\bar{\psi}_{-}],
~~
\hat{\psi} =\left[ \begin{array}{c} \psi_{+} \\ \psi_{-} \end{array}\right]
\end{align}
were introduced in the first step,
and where we performed the Larkin-Ovchinnikov rotation with the help of a matrix
\begin{align}
\hat{L} = \frac{1}{\sqrt{2}}\left[ \begin{array}{cc} 1 & - 1 \\ 1 & 1\end{array} \right], ~~
\hat{L}^{-1} = \frac{1}{\sqrt{2}}\left[ \begin{array}{cc} 1 & 1 \\ -1 & 1\end{array} \right],
\end{align}
in the next step, 
and as a consequence introduced rotated electron fields
\begin{align}
\hat{\bar{\Psi}} = \hat{\bar{\psi}}\hat{L}^{-1} = [\bar{\Psi}^{\mathrm{q}},~\bar{\Psi}^{\mathrm{cl}}], ~~ \hat{\Psi} =\hat{L}\hat{\sigma}_{3}\hat{\psi} = \left[\begin{array}{c} \Psi^{\mathrm{cl}} \\ \Psi^{\mathrm{q}} \end{array} \right],
\end{align}
where $\bar{\Psi}^{\mathrm{q}} = \frac{1}{\sqrt{2}}(\bar{\Psi}^{+} - \bar{\Psi}^{-})$ and $\bar{\Psi}^{\mathrm{cl}} = \frac{1}{\sqrt{2}}(\bar{\Psi}^{+} + \bar{\Psi}^{-})$, and the same for the fields without the bars.
The Hamiltonian gets rotated as
\begin{align}
\hat{L}
\left[\begin{array}{cc} \theta_{+} & 0 \\ 0 & \theta_{-}\end{array}\right]\hat{L}^{-1}
=
\frac{1}{2} \left[\begin{array}{cc} \theta_{+} + \theta_{-} & \theta_{+} - \theta_{-} \\ \theta_{+}-\theta_{-} & \theta_{+}+\theta_{-}\end{array} \right] \equiv
\left[\begin{array}{cc} \theta^{\mathrm{cl}}& \theta^{\mathrm{q}} \\ \theta^{\mathrm{q}} & \theta^{\mathrm{cl}}\end{array} \right]. 
\end{align}
For example, if $\theta(x) = {\bf P}_{0}\cdot{\bf P}(x)$, we have
\begin{align}
\left[\begin{array}{cc} \theta^{\mathrm{cl}}& \theta^{\mathrm{q}} \\ \theta^{\mathrm{q}} & \theta^{\mathrm{cl}}\end{array} \right] =
\left[\begin{array}{cc} 
{\bf P}_{0}\cdot{\bf P}^{\mathrm{cl}} & 
{\bf P}_{0}\cdot{\bf P}^{\mathrm{q}}  \\ 
{\bf P}_{0}\cdot{\bf P}^{\mathrm{q}}  & 
{\bf P}_{0}\cdot{\bf P}^{\mathrm{cl}}
\end{array} \right]. 
\end{align}
If now the field $\theta$ is a composite field, such as $\theta = {\bf P}\cdot{\bf P}$, then we get for the Hamiltonian
\begin{align}
\left[\begin{array}{cc} \theta^{\mathrm{cl}}& \theta^{\mathrm{q}} \\ \theta^{\mathrm{q}} & \theta^{\mathrm{cl}}\end{array} \right] = \frac{1}{4}
\left[\begin{array}{cc} 
{\bf P}^{\mathrm{cl}}\cdot{\bf P}^{\mathrm{cl}} + {\bf P}^{\mathrm{q}}\cdot{\bf P}^{\mathrm{q}} & 
{\bf P}^{\mathrm{cl}}\cdot{\bf P}^{\mathrm{q}} + {\bf P}^{\mathrm{q}}\cdot{\bf P}^{\mathrm{cl}} \\ 
{\bf P}^{\mathrm{cl}}\cdot{\bf P}^{\mathrm{q}} + {\bf P}^{\mathrm{q}}\cdot{\bf P}^{\mathrm{cl}} & 
{\bf P}^{\mathrm{cl}}\cdot{\bf P}^{\mathrm{cl}} + {\bf P}^{\mathrm{q}}\cdot{\bf P}^{\mathrm{q}}
\end{array} \right]. 
\end{align}

We will be needing correlators of ${\bf P}$ fields when deriving the effective interaction between electrons. 
They are
\begin{align}
\langle P^{\mathrm{cl}}_{\alpha}(x_{1})P^{\mathrm{cl}}_{\beta}(x_{2}) \rangle
&
=
\frac{i}{V}\sum_{\epsilon, {\bf q}}\left( \delta_{\alpha\beta} - \frac{q_{\alpha}q_{\beta}}{q^2}\right)
A_{{\bf q}}^2
\left[
e^{i{\bf q}({\bf r}_{1}-{\bf r}_{2})-i\epsilon(t_{1}-t_{2})}
+
e^{-i{\bf q}({\bf r}_{1}-{\bf r}_{2})+i\epsilon(t_{1}-t_{2})}  
\right]
d^{\mathrm{K}}(\epsilon,{\bf q}),
\\
\langle P^{\mathrm{cl}}_{\alpha}(x_{1})P^{\mathrm{q}}_{\beta}(x_{2}) \rangle
&
=\frac{i}{V}\sum_{\epsilon, {\bf q}}\left( \delta_{\alpha\beta} - \frac{q_{\alpha}q_{\beta}}{q^2}\right)
A_{{\bf q}}^2
\left[
e^{i{\bf q}({\bf r}_{1}-{\bf r}_{2})-i\epsilon(t_{1}-t_{2})}
d^{\mathrm{R}}(\epsilon,{\bf q})
+
e^{-i{\bf q}({\bf r}_{1}-{\bf r}_{2})+i\epsilon(t_{1}-t_{2})}  
d^{\mathrm{A}}(\epsilon,{\bf q})
\right],
\\
\langle P^{\mathrm{q}}_{\alpha}(x_{1})P^{\mathrm{q}}_{\beta}(x_{2}) \rangle &=0.
\end{align}
Here the phonon Green functions read
\begin{align}
&
d^{\mathrm{R}/\mathrm{A}}(\epsilon,{\bf q}) = \frac{1}{\epsilon \pm i0 -\omega_{{\bf q}}},
\\
&
d^{\mathrm{K}}(\epsilon,{\bf q}) = {\cal F}^{\mathrm{B}}_{\epsilon}\left[ d^{\mathrm{R}}(\epsilon,{\bf q}) - d^{\mathrm{A}}(\epsilon,{\bf q}) \right],
\end{align}
where ${\cal F}^{\mathrm{B}}_{\epsilon} = \coth\left( \frac{\epsilon}{2T}\right)$ is the boson distribution function. 
We can simplify the correlators by $\epsilon \rightarrow -\epsilon$ and ${\bf q} \rightarrow -{\bf q}$ under the sum in corresponding terms,
\begin{align}
\langle P^{\mathrm{cl}}_{\alpha}(x_{1})P^{\mathrm{cl}}_{\beta}(x_{2}) \rangle
&
=
\frac{i}{V}\sum_{\epsilon, {\bf q}}\left( \delta_{\alpha\beta} - \frac{q_{\alpha}q_{\beta}}{q^2}\right)
A_{{\bf q}}^2
e^{i{\bf q}({\bf r}_{1}-{\bf r}_{2})-i\epsilon(t_{1}-t_{2})}
D^{\mathrm{K}}(\epsilon,{\bf q}) ,
\\
\langle P^{\mathrm{cl}}_{\alpha}(x_{1})P^{\mathrm{q}}_{\beta}(x_{2}) \rangle
&
=
\frac{i}{V}\sum_{\epsilon, {\bf q}}\left( \delta_{\alpha\beta} - \frac{q_{\alpha}q_{\beta}}{q^2}\right)
A_{{\bf q}}^2
e^{i{\bf q}({\bf r}_{1}-{\bf r}_{2})-i\epsilon(t_{1}-t_{2})}
D^{\mathrm{R}}(\epsilon,{\bf q}),
\\
\langle P^{\mathrm{q}}_{\alpha}(x_{1})P^{\mathrm{cl}}_{\beta}(x_{2}) \rangle
&
=
\frac{i}{V}\sum_{\epsilon, {\bf q}}\left( \delta_{\alpha\beta} - \frac{q_{\alpha}q_{\beta}}{q^2}\right)
A_{{\bf q}}^2
e^{i{\bf q}({\bf r}_{1}-{\bf r}_{2})-i\epsilon(t_{1}-t_{2})}
D^{\mathrm{A}}(\epsilon,{\bf q}),
\end{align}
where now updated phonon Green functions read as
\begin{align}
&
D^{\mathrm{R}/\mathrm{A}}(\epsilon,{\bf q}) = \frac{2\omega_{{\bf q}}}{(\epsilon \pm i0 )^2 - \omega_{{\bf q}}^2},
\\
&
D^{\mathrm{K}}(\epsilon,{\bf q}) = {\cal F}^{\mathrm{B}}_{\epsilon}\left[ D^{\mathrm{R}}(\epsilon,{\bf q})  -  D^{\mathrm{A}}(\epsilon,{\bf q}) \right].
\end{align}
It is also worth noticing that the displacement fields commute
\begin{align}
P^{\mathrm{cl}}_{\alpha}(x_{1})P^{\mathrm{cl}}_{\beta}(x_{2}) 
=
P^{\mathrm{cl}}_{\beta}(x_{2})P^{\mathrm{cl}}_{\alpha}(x_{1}) .
\end{align}
Finally, 
\begin{align}
\frac{1}{V}\sum_{\bf q}(...) \rightarrow \int_{\bf q}(...)
\end{align}
will be made.

%---------------------------------------------------------------------------------------
\subsection{Effective interaction between electrons: one-phonon interaction}
%---------------------------------------------------------------------------------------
Interaction between electrons with two phonons is 
\begin{align}
H_{\mathrm{e}-2\mathrm{ph}} 
= g\int_{{\bf r}} \psi^{\dag}(x)\psi(x) [{\bf P}_{0}+{\bf P}(x)]\cdot [{\bf P}_{0}+{\bf P}(x)]
= g\int_{{\bf r}} \psi^{\dag}(x)\psi(x) [{\bf P}_{0}^2+{\bf P}^2(x) + 2{\bf P}_{0}\cdot{\bf P}(x)] .
\end{align} 
First term above, i.e. with ${\bf P}_{0}^2$, is just the chemical potential of electrons. 
Second term was studied in previous works and the results are understood.
Third term is new, and is a subject of the present study.
\begin{align}
H_{\mathrm{e}-2\mathrm{ph}} 
\rightarrow 2g\int_{{\bf r}} \psi^{\dag}(x)\psi(x) {\bf P}_{0}\cdot{\bf P}(x) .
\end{align} 
Effective interaction between electrons is obtained by integrating the phonons out. 
We assume that ${\bf P}_{0} = P_{0}{\bf e}_{\alpha}$, i.e. the ferroelectric spontaneous polarization is pointing in $\alpha$ direction.
To the second order in electron-phonon interaction we get
\begin{align}
&
iS^{(2)}=
\frac{1}{2}\langle \left(iS_{\mathrm{e}-2\mathrm{ph}}\right) \left(iS_{\mathrm{e}-2\mathrm{ph}}\right) \rangle_{\mathrm{phonons}}
\\
=
&
-i 2P_{0}^2 g^2
\int_{\epsilon_{1},\epsilon_{2},\omega}
\int_{{\bf k}_{1},{\bf k}_{2},{\bf q}}
\left[\hat{\bar{\Psi}}(\epsilon_{1}+\omega;{\bf k}_{1}+{\bf q})\hat{\sigma}_{0}\hat{\Psi}(\epsilon_{1};{\bf k}_{1}) \right]
\left[\hat{\bar{\Psi}}(\epsilon_{2}-\omega;{\bf k}_{2}-{\bf q})\hat{\sigma}_{0}\hat{\Psi}(\epsilon_{2};{\bf k}_{2}) \right]
\left( 1- \frac{q_{\alpha}^2}{q^2}\right)A_{{\bf q}}^2 D^{\mathrm{K}}(\omega;{\bf q})
\\
&
-i 2P_{0}^2 g^2
\int_{\epsilon_{1},\epsilon_{2},\omega}
\int_{{\bf k}_{1},{\bf k}_{2},{\bf q}}
\left[\hat{\bar{\Psi}}(\epsilon_{1}+\omega;{\bf k}_{1}+{\bf q})\hat{\sigma}_{0}\hat{\Psi}(\epsilon_{1};{\bf k}_{1}) \right]
\left[\hat{\bar{\Psi}}(\epsilon_{2}-\omega;{\bf k}_{2}-{\bf q})\hat{\sigma}_{1}\hat{\Psi}(\epsilon_{2};{\bf k}_{2}) \right]
\left( 1- \frac{q_{\alpha}^2}{q^2}\right)A_{{\bf q}}^2 D^{\mathrm{R}}(\omega;{\bf q})
\\
&
-i 2P_{0}^2 g^2
\int_{\epsilon_{1},\epsilon_{2},\omega}
\int_{{\bf k}_{1},{\bf k}_{2},{\bf q}}
\left[\hat{\bar{\Psi}}(\epsilon_{1}+\omega;{\bf k}_{1}+{\bf q})\hat{\sigma}_{1}\hat{\Psi}(\epsilon_{1};{\bf k}_{1}) \right]
\left[\hat{\bar{\Psi}}(\epsilon_{2}-\omega;{\bf k}_{2}-{\bf q})\hat{\sigma}_{0}\hat{\Psi}(\epsilon_{2};{\bf k}_{2}) \right]
\left( 1- \frac{q_{\alpha}^2}{q^2}\right)A_{{\bf q}}^2 D^{\mathrm{A}}(\omega;{\bf q}).
\end{align}

%--------------------------------------------------------
\subsection{Cooper ladder}
%--------------------------------------------------------
Here we derive a prescription how to deal with the Cooper channel. Although, it is a standard procedure and the resulting equations are standard textbook, we wish to derive it from scratch.
General static four electron interaction relevant for the problem at hand after regrouping of the Grassmann fields for the Cooper channel appears to be
\begin{align}
iS_{\mathrm{int}} 
=
&
-i\int_{\{\epsilon\};\{{\bf k}\}}
V({\bf k}_{1}-{\bf k}_{2})
\left[
(\hat{\bar{\Psi}}_{1;\alpha}\hat{\sigma}_{0}\hat{\Psi}_{2;\alpha})
(\hat{\bar{\Psi}}_{3;\beta}\hat{\sigma}_{1}\hat{\Psi}_{4;\beta})
+
(\hat{\bar{\Psi}}_{1;\alpha}\hat{\sigma}_{1}\hat{\Psi}_{2;\alpha})
(\hat{\bar{\Psi}}_{3;\beta}\hat{\sigma}_{0}\hat{\Psi}_{4;\beta})
\right]
\delta_{q_{1}-q_{2},q_{4}-q_{3}}
\\
=
&
-i\int_{\{\epsilon\};\{{\bf k}\}}
V({\bf k}_{1}-{\bf k}_{2})
\left( \bar{\Psi}^{\mathrm{q}}_{1;\alpha}\bar{\Psi}^{\mathrm{cl}}_{3;\beta}  
+ 
\bar{\Psi}^{\mathrm{cl}}_{1;\alpha}\bar{\Psi}^{\mathrm{q}}_{3;\beta} \right)
\left( \Psi^{\mathrm{cl}}_{4;\beta}\Psi^{\mathrm{cl}}_{2;\alpha}  
+ 
\Psi^{\mathrm{q}}_{4;\beta}\Psi^{\mathrm{q}}_{2;\alpha} \right)
\delta_{q_{1}-q_{2},q_{4}-q_{3}}
\\
&
-i\int_{\{\epsilon\};\{{\bf k}\}}
V({\bf k}_{1}-{\bf k}_{2})
\left( \bar{\Psi}^{\mathrm{q}}_{1;\alpha}\bar{\Psi}^{\mathrm{q}}_{3;\beta}  
+ 
\bar{\Psi}^{\mathrm{cl}}_{1;\alpha}\bar{\Psi}^{\mathrm{cl}}_{3;\beta} \right)
\left( \Psi^{\mathrm{q}}_{4;\beta}\Psi^{\mathrm{cl}}_{2;\alpha}  
+ 
\Psi^{\mathrm{cl}}_{4;\beta}\Psi^{\mathrm{q}}_{2;\alpha} \right)
\delta_{q_{1}-q_{2},q_{4}-q_{3}},
\end{align}
where $\alpha$ and $\beta$ are spins, and where $q_{i} = (\epsilon_{i},{\bf k}_{i})$.
Under the sign of integral we replace ${\bf k}_{3/4} \rightarrow -{\bf k}_{3/4}$ and $\epsilon_{3/4} \rightarrow -\epsilon_{3/4}$ so as to meet the structure of the Cooper channel, and rewrite the interaction
\begin{align}
iS_{\mathrm{int}}=
&
-i\int_{\{\epsilon\};\{{\bf k}\}}
V({\bf k}_{1}-{\bf k}_{2})
\left( \bar{\Psi}^{\mathrm{q}}_{1;\alpha}\bar{\Psi}^{\mathrm{cl}}_{-3;\beta}  
+ 
\bar{\Psi}^{\mathrm{cl}}_{1;\alpha}\bar{\Psi}^{\mathrm{q}}_{-3;\beta} \right)
\left( \Psi^{\mathrm{cl}}_{-4;\beta}\Psi^{\mathrm{cl}}_{2;\alpha}  
+ 
\Psi^{\mathrm{q}}_{-4;\beta}\Psi^{\mathrm{q}}_{2;\alpha} \right)
\delta_{q_{1}-q_{3},q_{2}-q_{4}}
\\
&
-
i
\int_{\{\epsilon\};\{{\bf k}\}}
V({\bf k}_{1}-{\bf k}_{2})
\left( \bar{\Psi}^{\mathrm{q}}_{1;\alpha}\bar{\Psi}^{\mathrm{q}}_{-3;\beta}  
+ 
\bar{\Psi}^{\mathrm{cl}}_{1;\alpha}\bar{\Psi}^{\mathrm{cl}}_{-3;\beta} \right)
\left( \Psi^{\mathrm{q}}_{-4;\beta}\Psi^{\mathrm{cl}}_{2;\alpha}  
+ 
\Psi^{\mathrm{cl}}_{-4;\beta}\Psi^{\mathrm{q}}_{2;\alpha} \right)
\delta_{q_{1}-q_{3},q_{2}-q_{4}},
\end{align}
where attraction corresponds to $V({\bf q})<0$.

Let us now iterate the interaction and construct a Cooper ladder. 
We first relabel the indeces
\begin{align}
iS_{\mathrm{int}}=
&
-i\int_{\{\epsilon\};\{{\bf k}\}}
V({\bf k}_{1}-{\bf k}_{4})
\left( \bar{\Psi}^{\mathrm{q}}_{1;\alpha}\bar{\Psi}^{\mathrm{cl}}_{-2;\beta}  
+ 
\bar{\Psi}^{\mathrm{cl}}_{1;\alpha}\bar{\Psi}^{\mathrm{q}}_{-2;\beta} \right)
\left( \Psi^{\mathrm{cl}}_{-3;\beta}\Psi^{\mathrm{cl}}_{4;\alpha}  
+ 
\Psi^{\mathrm{q}}_{-3;\beta}\Psi^{\mathrm{q}}_{4;\alpha} \right)
\delta_{q_{1}-q_{2},q_{4}-q_{3}}
\\
&
-
i
\int_{\{\epsilon\};\{{\bf k}\}}
V({\bf k}_{1}-{\bf k}_{4})
\left( \bar{\Psi}^{\mathrm{q}}_{1;\alpha}\bar{\Psi}^{\mathrm{q}}_{-2;\beta}  
+ 
\bar{\Psi}^{\mathrm{cl}}_{1;\alpha}\bar{\Psi}^{\mathrm{cl}}_{-2;\beta} \right)
\left( \Psi^{\mathrm{q}}_{-3;\beta}\Psi^{\mathrm{cl}}_{4;\alpha}  
+ 
\Psi^{\mathrm{cl}}_{-3;\beta}\Psi^{\mathrm{q}}_{4;\alpha} \right)
\delta_{q_{1}-q_{2},q_{4}-q_{3}}.
\end{align}
This is done for our convenience and for further construction of the ladder. 
To second order in interaction,
\begin{align}
&
\frac{1}{2} \langle \left( iS_{\mathrm{int}} \right) \left( iS_{\mathrm{int}} \right) \rangle
\\
=
&
\frac{1}{2} \int_{\{ \epsilon\}}\int_{\{{\bf k} \}} V({\bf k}_{1} - {\bf k}_{5}) V({\bf k}_{5} - {\bf k}_{4})
\delta_{q_{1}-q_{2},q_{4}-q_{3}}
\\
&
\times
\left[ 
G^{\mathrm{R}}(-\epsilon_{5} + \epsilon_{4} - \epsilon_{3};-{\bf k}_{5} + {\bf k}_{4} - {\bf k}_{3})G^{\mathrm{K}}(\epsilon_{5};{\bf k}_{5})  
+
G^{\mathrm{K}}(-\epsilon_{5} + \epsilon_{4} - \epsilon_{3};-{\bf k}_{5} + {\bf k}_{4} - {\bf k}_{3})G^{\mathrm{R}}(\epsilon_{5};{\bf k}_{5})  
 \right]
 \\
 &
\times\left[ \left( \bar{\Psi}^{\mathrm{q}}_{1;\alpha}\bar{\Psi}^{\mathrm{cl}}_{-2;\beta}  
+ 
\bar{\Psi}^{\mathrm{cl}}_{1;\alpha}\bar{\Psi}^{\mathrm{q}}_{-2;\beta} \right)
\left( \Psi^{\mathrm{cl}}_{-3;\beta}\Psi^{\mathrm{cl}}_{4;\alpha}  
+ 
\Psi^{\mathrm{q}}_{-3;\beta}\Psi^{\mathrm{q}}_{4;\alpha} \right)
+
\left( \bar{\Psi}^{\mathrm{q}}_{1;\alpha}\bar{\Psi}^{\mathrm{q}}_{-2;\beta}  
+ 
\bar{\Psi}^{\mathrm{cl}}_{1;\alpha}\bar{\Psi}^{\mathrm{cl}}_{-2;\beta} \right)
\left( \Psi^{\mathrm{q}}_{-3;\beta}\Psi^{\mathrm{cl}}_{4;\alpha}  
+ 
\Psi^{\mathrm{cl}}_{-3;\beta}\Psi^{\mathrm{q}}_{4;\alpha} \right)
\right]
\end{align}
where we have relabelled the indeces for our convenience.
Recall that $G^{\mathrm{K}}(\epsilon;{\bf k}) = {\cal F}^{\mathrm{F}}_{\epsilon}\left(G^{\mathrm{R}}(\epsilon;{\bf k}) - G^{\mathrm{A}}(\epsilon;{\bf k}) \right) = -2\pi i {\cal F}^{\mathrm{F}}_{\epsilon} \delta(\epsilon - \xi_{\bf k})$ where ${\cal F}_{\epsilon}^{\mathrm{F}} = \tanh\left(\frac{\epsilon}{2T} \right)$ is the electron distribution function in equilibrium. Integral over the $\epsilon_{5}$ frequency is
\begin{align}
&
\int_{\epsilon_{5}}
\left[ 
G^{\mathrm{R}}(-\epsilon_{5} + \epsilon_{4} - \epsilon_{3};-{\bf k}_{5} + {\bf k}_{4} - {\bf k}_{3})G^{\mathrm{K}}(\epsilon_{5};{\bf k}_{5})  
+
G^{\mathrm{K}}(-\epsilon_{5} + \epsilon_{4} - \epsilon_{3};-{\bf k}_{5} + {\bf k}_{4} - {\bf k}_{3})G^{\mathrm{R}}(\epsilon_{5};{\bf k}_{5})  
 \right]
 \\
 &
 =
 -i 
 \frac{{\cal F}^{\mathrm{F}}_{\xi_{{\bf k}_{5} }} + {\cal F}^{\mathrm{F}}_{\xi_{-{\bf k}_{5} + {\bf k}_{4} -{\bf k}_{3} }} }{\epsilon_{4} - \epsilon_{3} -\xi_{{\bf k}_{5}} - \xi_{-{\bf k}_{5} + {\bf k}_{4} -{\bf k}_{3} } + i0}
 \rightarrow
  i 
 \frac{{\cal F}^{\mathrm{F}}_{\xi_{{\bf k}_{5} }} + {\cal F}^{\mathrm{F}}_{\xi_{-{\bf k}_{5} + {\bf k}_{4} -{\bf k}_{3} }} }{\xi_{{\bf k}_{5}} + \xi_{-{\bf k}_{5} + {\bf k}_{4} -{\bf k}_{3} }- i0},
\end{align}
where static limit was taken in the transformation under the right arrow.
In our case the static part of the interaction is 
\begin{align}
V({\bf q}) 
= 2g^2P_{0}^2 \left( 1- \frac{q_{\alpha}^2}{q^2}\right)A_{{\bf q}}^2 D^{\mathrm{R}}(\omega =0;{\bf q}) 
 = -\frac{g^2 P_{0}^2}{\pi}\left( 1- \frac{q_{\alpha}^2}{q^2}\right) \frac{\Omega_{0}^2}{\omega_{\bf q}^2},
\end{align}
recall that $A_{{\bf q}}^2 \approx \frac{1}{4\pi}\frac{\Omega_{0}^2}{\omega_{{\bf q}}} $. It is clear that the interaction between electrons is attractive.
The Cooper ladder then reads 
\begin{align}
\Gamma({\bf k}_{1}-{\bf k}_{4}) = V({\bf k}_{1}-{\bf k}_{4}) 
- \frac{1}{2}\int_{\bf p} V({\bf k}_{1}-{\bf p}) \frac{{\cal F}^{\mathrm{F}}_{\xi_{{\bf p} }} + {\cal F}^{\mathrm{F}}_{\xi_{-{\bf p} + {\bf k}_{4} -{\bf k}_{3} }} }{\xi_{{\bf p}} + \xi_{-{\bf p} + {\bf k}_{4} -{\bf k}_{3} }} \Gamma({\bf p}-{\bf k}_{4}),
\end{align}
assuming that ${\bf k}_{4} = {\bf k}_{3} $, we get
\begin{align}
\Gamma({\bf k}_{1}-{\bf k}_{4}) = V({\bf k}_{1}-{\bf k}_{4}) 
- 
\frac{1}{2}
\int_{\bf p} V({\bf k}_{1}-{\bf p}) 
\frac{{\cal F}^{\mathrm{F}}_{\xi_{{\bf p} }}  }{\xi_{{\bf p}} } \Gamma({\bf p}-{\bf k}_{4}).
\end{align}
It is convenient to apply a change of variables from ${\bf p}$ to $Q=\vert {\bf p} - {\bf k}_{1}\vert$ and $p$ as
\begin{align}\label{transform}
\int \frac{d^{3}p}{(2\pi)^3}(..) = \frac{1}{k_{1}}\int_{0}^{\infty} \frac{pdp}{2\pi^2 } \int_{\vert p-k_{1} \vert}^{{\vert p+k_{1} \vert}} \frac{QdQ}{2} (..).
\end{align}
For example, such a trick is given in Ref. [2]. We put $\left( 1- \frac{q_{\alpha}^2}{q^2}\right) \rightarrow \frac{2}{3}$ in the interaction, so it becomes $V({\bf q})  = -\frac{2g^2 P_{0}^2}{3\pi V}\frac{\Omega_{0}^2}{\omega_{\bf q}^2}$. 
Note that due to the anisotropic interaction, the superconductivity order parameter is expected to be anisotropic as well. This is a question for future research.
We assume that $\Gamma({\bf p}-{\bf k}_{4})$ is momentum independent, estimated at the Fermi momentum. Then, after the transform Eq. (\ref{transform}), we get for the integral over $Q$
\begin{align}
\int_{\vert p-k_{1} \vert}^{\vert p+k_{1} \vert} \frac{QdQ}{2}V(Q)
=
-\frac{2g^2 P_{0}^2 \Omega_{0}^2}{3\pi } \int_{\vert p-k_{1} \vert}^{\vert p+k_{1} \vert} \frac{QdQ}{2}\frac{1}{\omega_{Q}^2}
=
-\frac{g^2 P_{0}^2 \Omega_{0}^2}{6\pi  s^2} 
\ln\left[ \frac{\omega_{\mathrm{TO}}^2+ s^2\vert p+k_{1} \vert^2}{\omega_{\mathrm{TO}}^2+ s^2\vert p-k_{1} \vert^2}\right]
\approx
-\frac{g^2 P_{0}^2 \Omega_{0}^2}{6\pi  s^2} 
\ln\left[ \frac{\omega_{\mathrm{TO}}^2+ T_{\mathrm{BG}}^2}{\omega_{\mathrm{TO}}^2+ \frac{s^2}{v_{\mathrm{F}}^2}\xi_{\bf p}^2} \right],
\end{align}
where $T_{\mathrm{BG}} = s2p_{\mathrm{F}}$ is the Bloch-Gruneisen frequency. If $v_{\mathrm{F}} \gg s$, then the $\frac{s^2}{v_{\mathrm{F}}^2}\xi_{\bf p}^2$ term can be dropped as compared with $\omega_{\mathrm{TO}}^2$. Remaining integral is
\begin{align}
\int_{0}^{\infty}\frac{pdp}{2\pi^2} \frac{{\cal F}^{\mathrm{F}}_{\xi_{{\bf p} }}  }{\xi_{{\bf p}} }
=
\frac{m}{4\pi^2}
\int_{-\mu}^{\omega_{\mathrm{D}}}d\xi_{\bf p} \frac{{\cal F}^{\mathrm{F}}_{\xi_{{\bf p} }}  }{\xi_{{\bf p}} }
=
\frac{m}{4\pi^2}
\left(\int_{0}^{\mu} + \int_{0}^{\mu} \right)  d\xi_{\bf p} \frac{{\cal F}^{\mathrm{F}}_{\xi_{{\bf p} }}  }{\xi_{{\bf p}} }
=
\frac{m}{2\pi^2}\ln\left( \frac{\mu }{T} \right),
\end{align}
valid when $\omega_{\mathrm{D}}\gg\mu$, which is the case in STO.
It is important that both upper limits are the same.
Overall, we have
\begin{align}
\int_{\bf p} V({\bf k}_{1}-{\bf p}) 
\frac{{\cal F}^{\mathrm{F}}_{\xi_{{\bf p} }}  }{\xi_{{\bf p}} }
=
-\frac{g^2 m P_{0}^2 \Omega_{0}^2}{12 \pi^3  s^2 p_{\mathrm{F}}} 
\ln\left[ \frac{\omega_{\mathrm{TO}}^2+ T_{\mathrm{BG}}^2}{\omega_{\mathrm{TO}}^2} \right]
\ln\left( \frac{\mu }{T} \right).
\end{align}
Superconducting transition temperature is 
\begin{align}
T_{\mathrm{c}}^{\mathrm{SC}} = \mu e^{-\frac{\nu}{G}},
\end{align}
where $\nu = \frac{m p_{F}}{2\pi^2}$ is the density of states, and
\begin{align}
G = \frac{g^2 m^2 P_{0}^2 \Omega_{0}^2}{24 \pi^5 s^2 } 
\ln\left[ \frac{\omega_{\mathrm{TO}}^2+ T_{\mathrm{BG}}^2}{\omega_{\mathrm{TO}}^2} \right]
\end{align}
is the effective interaction. There is a maximum in $T_{\mathrm{c}}^{\mathrm{SC}}$ at some value of $\mu$.

%----------------------------------------------------------------------------------------------------------------------------
\subsection{Self-energy due to one-phonon interaction: imaginary part}
%----------------------------------------------------------------------------------------------------------------------------
To derive kinetic equation, we need to calculate self-energy due to electron-phonon interaction.
We contract the interaction
\begin{align}
&
-i 2P_{0}^2 g^2
\int_{\epsilon_{1},\epsilon_{2},\omega}
\int_{{\bf k}_{1},{\bf k}_{2},{\bf q}}
\left\langle
\left[\hat{\bar{\Psi}}(\epsilon_{1}+\omega;{\bf k}_{1}+{\bf q})\hat{\sigma}_{0}\hat{\Psi}(\epsilon_{1};{\bf k}_{1}) \right]
\left[\hat{\bar{\Psi}}(\epsilon_{2}-\omega;{\bf k}_{2}-{\bf q})\hat{\sigma}_{0}\hat{\Psi}(\epsilon_{2};{\bf k}_{2}) \right]
\right\rangle
\left( 1- \frac{q_{\alpha}^2}{q^2}\right)A_{{\bf q}}^2 D^{\mathrm{K}}(\omega;{\bf q})
\\
&
-i 2P_{0}^2 g^2
\int_{\epsilon_{1},\epsilon_{2},\omega}
\int_{{\bf k}_{1},{\bf k}_{2},{\bf q}}
\left\langle
\left[\hat{\bar{\Psi}}(\epsilon_{1}+\omega;{\bf k}_{1}+{\bf q})\hat{\sigma}_{0}\hat{\Psi}(\epsilon_{1};{\bf k}_{1}) \right]
\left[\hat{\bar{\Psi}}(\epsilon_{2}-\omega;{\bf k}_{2} - {\bf q})\hat{\sigma}_{1}\hat{\Psi}(\epsilon_{2};{\bf k}_{2}) \right]
\right\rangle
\left( 1- \frac{q_{\alpha}^2}{q^2}\right)A_{{\bf q}}^2 D^{\mathrm{R}}(\omega;{\bf q})
\\
&
-i 2P_{0}^2 g^2
\int_{\epsilon_{1},\epsilon_{2},\omega}
\int_{{\bf k}_{1},{\bf k}_{2},{\bf q}}
\left\langle
\left[\hat{\bar{\Psi}}(\epsilon_{1}+\omega;{\bf k}_{1}+{\bf q})\hat{\sigma}_{1}\hat{\Psi}(\epsilon_{1};{\bf k}_{1}) \right]
\left[\hat{\bar{\Psi}}(\epsilon_{2}-\omega;{\bf k}_{2}-{\bf q})\hat{\sigma}_{0}\hat{\Psi}(\epsilon_{2};{\bf k}_{2}) \right]
\right\rangle
\left( 1- \frac{q_{\alpha}^2}{q^2}\right)A_{{\bf q}}^2 D^{\mathrm{A}}(\omega;{\bf q}),
\end{align}
where the contractions read as
\begin{align}
&
\int_{\epsilon_{1},\epsilon_{2},\omega}
\left\langle
\left[\hat{\bar{\Psi}}(\epsilon_{1}+\omega;{\bf k}_{1}+{\bf q})\hat{\sigma}_{0}\hat{\Psi}(\epsilon_{1};{\bf k}_{1}) \right]
\left[\hat{\bar{\Psi}}(\epsilon_{2}-\omega;{\bf k}_{2}-{\bf q})\hat{\sigma}_{1}\hat{\Psi}(\epsilon_{2};{\bf k}_{2}) \right]
\right\rangle
\\
=
&
\int_{\epsilon_{1},\omega}
\bar{\Psi}^{\mathrm{q}}(\epsilon_{1};{\bf k}_{1}) 
\Psi^{\mathrm{cl}}(\epsilon_{1};{\bf k}_{1})
iG^{\mathrm{K}}(\epsilon_{1}-\omega;{\bf k}_{1}-{\bf q})
\\
+
&
\int_{\epsilon_{1},\omega}
\bar{\Psi}^{\mathrm{cl}}(\epsilon_{1};{\bf k}_{1}) 
\Psi^{\mathrm{q}}(\epsilon_{1};{\bf k}_{1})
iG^{\mathrm{K}}(\epsilon_{1}+\omega;{\bf k}_{1}+{\bf q})
\\
+
&
\int_{\epsilon_{1},\omega}
\bar{\Psi}^{\mathrm{cl}}(\epsilon_{1};{\bf k}_{1}) 
\Psi^{\mathrm{cl}}(\epsilon_{1};{\bf k}_{1})
\left[ 
iG^{\mathrm{R}}(\epsilon_{1}+\omega;{\bf k}_{1}+{\bf q}) 
+ 
iG^{\mathrm{A}}(\epsilon_{1}-\omega;{\bf k}_{1}-{\bf q}) 
\right]
\\
+
&
\int_{\epsilon_{1},\omega}
\bar{\Psi}^{\mathrm{q}}(\epsilon_{1};{\bf k}_{1}) 
\Psi^{\mathrm{q}}(\epsilon_{1};{\bf k}_{1})
\left[ 
iG^{\mathrm{R}}(\epsilon_{1}-\omega;{\bf k}_{1}-{\bf q}) 
+ 
iG^{\mathrm{A}}(\epsilon_{1}+\omega;{\bf k}_{1}+{\bf q}) 
\right],
\end{align}
where we have relabelled the indexes where needed.
Similarly
\begin{align}
&
\int_{\epsilon_{1},\epsilon_{2},\omega}
\left\langle
\left[\hat{\bar{\Psi}}(\epsilon_{1}+\omega;{\bf k}_{1}+{\bf q})\hat{\sigma}_{1}\hat{\Psi}(\epsilon_{1};{\bf k}_{1}) \right]
\left[\hat{\bar{\Psi}}(\epsilon_{2}-\omega;{\bf k}_{2}-{\bf q})\hat{\sigma}_{0}\hat{\Psi}(\epsilon_{2};{\bf k}_{2}) \right]
\right\rangle
\\
=
&
\int_{\epsilon_{1},\omega}
\bar{\Psi}^{\mathrm{q}}(\epsilon_{1};{\bf k}_{1}) 
\Psi^{\mathrm{cl}}(\epsilon_{1};{\bf k}_{1})
iG^{\mathrm{K}}(\epsilon_{1}+\omega;{\bf k}_{1}+{\bf q})
\\
+
&
\int_{\epsilon_{1},\omega}
\bar{\Psi}^{\mathrm{cl}}(\epsilon_{1};{\bf k}_{1}) 
\Psi^{\mathrm{q}}(\epsilon_{1};{\bf k}_{1})
iG^{\mathrm{K}}(\epsilon_{1}-\omega;{\bf k}_{1}-{\bf q})
\\
+
&
\int_{\epsilon_{1},\omega}
\bar{\Psi}^{\mathrm{cl}}(\epsilon_{1};{\bf k}_{1}) 
\Psi^{\mathrm{cl}}(\epsilon_{1};{\bf k}_{1})
\left[ 
iG^{\mathrm{R}}(\epsilon_{1}-\omega;{\bf k}_{1}-{\bf q}) 
+ 
iG^{\mathrm{A}}(\epsilon_{1}+\omega;{\bf k}_{1}+{\bf q}) 
\right]
\\
+
&
\int_{\epsilon_{1},\omega}
\bar{\Psi}^{\mathrm{q}}(\epsilon_{1};{\bf k}_{1}) 
\Psi^{\mathrm{q}}(\epsilon_{1};{\bf k}_{1})
\left[ 
iG^{\mathrm{R}}(\epsilon_{1}+\omega;{\bf k}_{1}+{\bf q}) 
+ 
iG^{\mathrm{A}}(\epsilon_{1}-\omega;{\bf k}_{1}-{\bf q}) 
\right].
\end{align}
Last contraction reads as
\begin{align}
&
\int_{\epsilon_{1},\epsilon_{2},\omega}
\left\langle
\left[\hat{\bar{\Psi}}(\epsilon_{1}+\omega;{\bf k}_{1}+{\bf q})\hat{\sigma}_{0}\hat{\Psi}(\epsilon_{1};{\bf k}_{1}) \right]
\left[\hat{\bar{\Psi}}(\epsilon_{2}-\omega;{\bf k}_{2}-{\bf q})\hat{\sigma}_{0}\hat{\Psi}(\epsilon_{2};{\bf k}_{2}) \right]
\right\rangle
\\
=
&
\int_{\epsilon_{1},\omega}
\bar{\Psi}^{\mathrm{q}}(\epsilon_{1};{\bf k}_{1}) 
\Psi^{\mathrm{cl}}(\epsilon_{1};{\bf k}_{1})
\left[
iG^{\mathrm{R}}(\epsilon_{1}+\omega;{\bf k}_{1}+{\bf q})
+
iG^{\mathrm{R}}(\epsilon_{1}-\omega;{\bf k}_{1}-{\bf q})
\right]
\\
+
&
\int_{\epsilon_{1},\omega}
\bar{\Psi}^{\mathrm{cl}}(\epsilon_{1};{\bf k}_{1}) 
\Psi^{\mathrm{q}}(\epsilon_{1};{\bf k}_{1})
\left[
iG^{\mathrm{A}}(\epsilon_{1}+\omega;{\bf k}_{1}+{\bf q})
+
iG^{\mathrm{A}}(\epsilon_{1}-\omega;{\bf k}_{1}-{\bf q})
\right]
\\
+
&
\int_{\epsilon_{1},\omega}
\bar{\Psi}^{\mathrm{q}}(\epsilon_{1};{\bf k}_{1}) 
\Psi^{\mathrm{q}}(\epsilon_{1};{\bf k}_{1})
\left[
iG^{\mathrm{K}}(\epsilon_{1}+\omega;{\bf k}_{1}+{\bf q})
+
iG^{\mathrm{K}}(\epsilon_{1}-\omega;{\bf k}_{1}-{\bf q})
\right].
\end{align}
Let us first write down a general expression for the self-energy.
\begin{align}
\langle S_{\mathrm{int}}\rangle 
=
&
 -i \int_{\epsilon;{\bf k}}
\bar{\Psi}^{\mathrm{cl}}(\epsilon;{\bf k}) 
\Psi^{\mathrm{cl}}(\epsilon;{\bf k}) 
\Sigma^{\mathrm{cl-cl}}(\epsilon;{\bf k})
 -i \int_{\epsilon;{\bf k}}
\bar{\Psi}^{\mathrm{q}}(\epsilon;{\bf k}) 
\Psi^{\mathrm{q}}(\epsilon;{\bf k}) 
\Sigma^{\mathrm{K}}(\epsilon;{\bf k})
\\
&
 -i \int_{\epsilon;{\bf k}}
\bar{\Psi}^{\mathrm{q}}(\epsilon;{\bf k}) 
\Psi^{\mathrm{cl}}(\epsilon;{\bf k}) 
\Sigma^{\mathrm{R}}(\epsilon;{\bf k})
 -i \int_{\epsilon;{\bf k}}
\bar{\Psi}^{\mathrm{cl}}(\epsilon;{\bf k}) 
\Psi^{\mathrm{q}}(\epsilon;{\bf k}) 
\Sigma^{\mathrm{A}}(\epsilon;{\bf k})
\\
=
&
 -i \int_{\epsilon;{\bf k}}
  \hat{\bar{\Psi}}(\epsilon;{\bf k})  
 \left[\begin{array}{cc} 
 \Sigma^{\mathrm{R}}(\epsilon;{\bf k}) & \Sigma^{\mathrm{K}}(\epsilon;{\bf k}) \\
 \Sigma^{\mathrm{cl-cl}}(\epsilon;{\bf k})  & \Sigma^{\mathrm{A}}(\epsilon;{\bf k})
 \end{array}\right]
 \hat{\Psi}(\epsilon;{\bf k}),
\end{align}
where, recall, the fields with a hat are the spinors in the Keldysh space,
\begin{align}
\hat{\bar{\Psi}} = [\bar{\Psi}^{\mathrm{q}},~\bar{\Psi}^{\mathrm{cl}}], ~~ \hat{\Psi} = \left[\begin{array}{c} \Psi^{\mathrm{cl}} \\ \Psi^{\mathrm{q}} \end{array} \right].
\end{align}
It is expected that $\Sigma^{\mathrm{cl-cl}}(\epsilon;{\bf k}) = 0$, which we will show below for our model.
The expressions for the components of the self-energy are 
\begin{align}
\Sigma^{\mathrm{cl-cl}}(\epsilon;{\bf k}) 
=
&
\int_{\omega;{\bf q}}
\left[ 
iG^{\mathrm{R}}(\epsilon+\omega;{\bf k}+{\bf q}) 
+ 
iG^{\mathrm{A}}(\epsilon-\omega;{\bf k}-{\bf q}) 
\right]U^{\mathrm{R}}(\omega;{\bf q})
\\
+
&
\int_{\omega;{\bf q}}
\left[ 
iG^{\mathrm{R}}(\epsilon-\omega;{\bf k}-{\bf q}) 
+ 
iG^{\mathrm{A}}(\epsilon+\omega;{\bf k}+{\bf q}) 
\right]U^{\mathrm{A}}(\omega;{\bf q}),
\end{align}
\begin{align}
\Sigma^{\mathrm{K}}(\epsilon;{\bf k})
=
&
\int_{\omega;{\bf q}}
\left[ 
iG^{\mathrm{R}}(\epsilon-\omega;{\bf k}-{\bf q}) 
+ 
iG^{\mathrm{A}}(\epsilon+\omega;{\bf k}+{\bf q}) 
\right]
U^{\mathrm{R}}(\omega;{\bf q})
\\
+
&
\int_{\omega;{\bf q}}
\left[ 
iG^{\mathrm{R}}(\epsilon+\omega;{\bf k}+{\bf q}) 
+ 
iG^{\mathrm{A}}(\epsilon-\omega;{\bf k}-{\bf q}) 
\right]
U^{\mathrm{A}}(\omega;{\bf q})
\\
+
&
\int_{\omega;{\bf q}}
\left[
iG^{\mathrm{K}}(\epsilon+\omega;{\bf k}+{\bf q})
+
iG^{\mathrm{K}}(\epsilon-\omega;{\bf k}-{\bf q})
\right]
U^{\mathrm{K}}(\omega;{\bf q}),
\end{align}

\begin{align}
\Sigma^{\mathrm{R}}(\epsilon;{\bf k}) 
=
&
\int_{\omega;{\bf q}}
\left[ 
iG^{\mathrm{K}}(\epsilon+\omega;{\bf k}+{\bf q}) 
U^{\mathrm{A}}(\omega;{\bf q})
+
iG^{\mathrm{K}}(\epsilon-\omega;{\bf k}-{\bf q}) 
U^{\mathrm{R}}(\omega;{\bf q})
\right]
\\
+
&
\int_{\omega;{\bf q}}
\left[
iG^{\mathrm{R}}(\epsilon+\omega;{\bf k}+{\bf q})
+
iG^{\mathrm{R}}(\epsilon-\omega;{\bf k}-{\bf q})
\right]U^{\mathrm{K}}(\omega;{\bf q}),
\end{align}

\begin{align}
\Sigma^{\mathrm{A}}(\epsilon;{\bf k}) 
=
&
\int_{\omega;{\bf q}}
\left[ 
iG^{\mathrm{K}}(\epsilon+\omega;{\bf k}+{\bf q}) 
U^{\mathrm{R}}(\omega;{\bf q})
+
iG^{\mathrm{K}}(\epsilon-\omega;{\bf k}-{\bf q}) 
U^{\mathrm{A}}(\omega;{\bf q})
\right]
\\
+
&
\int_{\omega;{\bf q}}
\left[
iG^{\mathrm{A}}(\epsilon+\omega;{\bf k}+{\bf q})
+
iG^{\mathrm{A}}(\epsilon-\omega;{\bf k}-{\bf q})
\right]U^{\mathrm{K}}(\omega;{\bf q}).
\end{align}

Let us demonstrate that $\Sigma^{\mathrm{cl-cl}}(\epsilon;{\bf k}) = 0$. 
We use $D^{\mathrm{A}}(-\omega;{\bf q}) = D^{\mathrm{R}}(\omega;{\bf q})$, which allows to double certain terms.
Furthermore, we notice that in the expression for the $\Sigma^{\mathrm{cl-cl}}(\epsilon;{\bf k}) $ all the residues are on the same half of the complex space plane. Thus, indeed, due to integration over the frequency $\omega$
\begin{align}
-i 4P_{0}^2 g^2
\int_{\epsilon;{\bf k}}
\bar{\Psi}^{\mathrm{cl}}(\epsilon;{\bf k}) 
\Psi^{\mathrm{cl}}(\epsilon;{\bf k})
\int_{\omega;{\bf q}}
\left[ 
iG^{\mathrm{R}}(\epsilon+\omega;{\bf k}+{\bf q}) 
+ 
iG^{\mathrm{A}}(\epsilon-\omega;{\bf k}-{\bf q}) 
\right]
\left(1-\frac{q_{\alpha}^2}{q^2}\right)A_{{\bf q}}^2D^{\mathrm{R}}(\omega;{\bf q}) =0.
\end{align}
The Keldysh part of the self-energy reads
\begin{align}
&
-i 4P_{0}^2 g^2
\int_{\epsilon;{\bf k}}
\bar{\Psi}^{\mathrm{q}}(\epsilon;{\bf k}) 
\Psi^{\mathrm{q}}(\epsilon;{\bf k})
\\
&
\times
\int_{\omega;{\bf q}}
\left\{
\left[ 
iG^{\mathrm{R}}(\epsilon-\omega;{\bf k}-{\bf q}) 
+ 
iG^{\mathrm{A}}(\epsilon+\omega;{\bf k}+{\bf q}) 
\right]
D^{\mathrm{R}}(\omega;{\bf q}) 
+
iG^{\mathrm{K}}(\epsilon+\omega;{\bf k}+{\bf q})
D^{\mathrm{K}}(\omega;{\bf q})
\right\}\left(1-\frac{q_{\alpha}^2}{q^2}\right)A_{{\bf q}}^2
\\
=
&
-i4 P_{0}^2 g^2 (2\pi i)
\int_{\epsilon;{\bf k}}
\bar{\Psi}^{\mathrm{q}}(\epsilon;{\bf k}) 
\Psi^{\mathrm{q}}(\epsilon;{\bf k})
\int_{\bf q}
\left[
{\cal F}^{\mathrm{F}}_{\xi_{{\bf k} + {\bf q}}} 
{\cal F}^{\mathrm{B}}_{\xi_{{\bf k} + {\bf q}}-\epsilon} 
-1
\right]
\left[
\delta\left( \epsilon - \xi_{{\bf k}+{\bf q}} -\omega_{\bf q}\right) 
-
\delta\left( \epsilon - \xi_{{\bf k}+{\bf q}} +\omega_{\bf q}\right) 
\right]
\left(1-\frac{q_{\alpha}^2}{q^2}\right)A_{{\bf q}}^2,
\end{align}
where
\begin{align}
&
\int_{\omega;{\bf q}}
\left[ 
iG^{\mathrm{R}}(\epsilon-\omega;{\bf k}-{\bf q}) 
+ 
iG^{\mathrm{A}}(\epsilon+\omega;{\bf k}+{\bf q}) 
\right]
\left(1-\frac{q_{\alpha}^2}{q^2}\right)A_{{\bf q}}^2
D^{\mathrm{R}}(\omega;{\bf q}) 
\\
=
&
-2\pi i \int_{\bf q}
\left[
\delta\left( \epsilon - \xi_{{\bf k}+{\bf q}} -\omega_{\bf q}\right) 
-
\delta\left( \epsilon - \xi_{{\bf k}+{\bf q}} +\omega_{\bf q}\right) 
\right] \left(1-\frac{q_{\alpha}^2}{q^2}\right)A_{{\bf q}}^2,
\end{align}
and
\begin{align}
&
\int_{\omega;{\bf q}}
iG^{\mathrm{K}}(\epsilon+\omega;{\bf k}+{\bf q})
D^{\mathrm{K}}(\omega;{\bf q})
\left(1-\frac{q_{\alpha}^2}{q^2}\right)A_{{\bf q}}^2
\\
=
&
2\pi i\int_{\bf q}
{\cal F}^{\mathrm{F}}_{\xi_{{\bf k} + {\bf q}}} 
{\cal F}^{\mathrm{B}}_{\xi_{{\bf k} + {\bf q}}-\epsilon} 
\left[
\delta\left( \epsilon - \xi_{{\bf k}+{\bf q}} -\omega_{\bf q}\right) 
-
\delta\left( \epsilon - \xi_{{\bf k}+{\bf q}} +\omega_{\bf q}\right) 
\right]
\left(1-\frac{q_{\alpha}^2}{q^2}\right)A_{{\bf q}}^2.
\end{align}
Another term after relabelling ceratin indexes is
\begin{align}
&
-i 4P_{0}^2 g^2
\int_{\epsilon;{\bf k}}
\bar{\Psi}^{\mathrm{cl}}(\epsilon;{\bf k}) 
\Psi^{\mathrm{q}}(\epsilon;{\bf k})
\int_{\omega;{\bf q}}
\left[ 
iG^{\mathrm{K}}(\epsilon+\omega;{\bf k}+{\bf q}) D^{\mathrm{R}}(\omega;{\bf q})
+ 
iG^{\mathrm{A}}(\epsilon+\omega;{\bf k}+{\bf q}) D^{\mathrm{K}}(\omega;{\bf q})
\right]
\left(1-\frac{q_{\alpha}^2}{q^2}\right)A_{{\bf q}}^2
\\
=
&
-i 4P_{0}^2 g^2
\int_{\epsilon;{\bf k}}
\bar{\Psi}^{\mathrm{cl}}(\epsilon;{\bf k}) 
\Psi^{\mathrm{q}}(\epsilon;{\bf k})
\int_{\bf q} 
\left[ 
{\cal F}^{\mathrm{F}}_{\xi_{{\bf k} + {\bf q}}} 
\frac{2\omega_{\bf q}}{\left( \epsilon - \xi_{{\bf k} + {\bf q}} + i0 \right)^2 - \omega_{\bf q}^2}
+
{\cal F}^{\mathrm{B}}_{\omega_{\bf q}} 
\frac{2\left( \epsilon - \xi_{{\bf k} + {\bf q}}  \right)}{\left( \epsilon - \xi_{{\bf k} + {\bf q}} + i0 \right)^2 - \omega_{\bf q}^2}
\right]
\left(1-\frac{q_{\alpha}^2}{q^2}\right)A_{{\bf q}}^2 ,
\end{align}
where the integral is
\begin{align}
&
\int_{\omega;{\bf q}}
\left[ 
iG^{\mathrm{K}}(\epsilon+\omega;{\bf k}+{\bf q}) D^{\mathrm{R}}(\omega;{\bf q})
+ 
iG^{\mathrm{A}}(\epsilon+\omega;{\bf k}+{\bf q}) D^{\mathrm{K}}(\omega;{\bf q})
\right]
\left(1-\frac{q_{\alpha}^2}{q^2}\right)A_{{\bf q}}^2
\\
=
& 
\int_{\bf q} 
\left[ 
{\cal F}^{\mathrm{F}}_{\xi_{{\bf k} + {\bf q}}} 
\frac{2\omega_{\bf q}}{\left( \epsilon - \xi_{{\bf k} + {\bf q}} + i0 \right)^2 - \omega_{\bf q}^2}
+
{\cal F}^{\mathrm{B}}_{\omega_{\bf q}} 
\frac{2\left( \epsilon - \xi_{{\bf k} + {\bf q}}  \right)}{\left( \epsilon - \xi_{{\bf k} + {\bf q}} + i0 \right)^2 - \omega_{\bf q}^2}
\right]
\left(1-\frac{q_{\alpha}^2}{q^2}\right)A_{{\bf q}}^2.
\end{align}
Finally, 
\begin{align}
&
-i 4P_{0}^2 g^2
\int_{\epsilon;{\bf k}}
\bar{\Psi}^{\mathrm{q}}(\epsilon;{\bf k}) 
\Psi^{\mathrm{cl}}(\epsilon;{\bf k})
\int_{\omega;{\bf q}}
\left[ 
iG^{\mathrm{K}}(\epsilon+\omega;{\bf k}+{\bf q}) D^{\mathrm{A}}(\omega;{\bf q})
+ 
iG^{\mathrm{R}}(\epsilon+\omega;{\bf k}+{\bf q}) D^{\mathrm{K}}(\omega;{\bf q})
\right]
\left(1-\frac{q_{\alpha}^2}{q^2}\right)A_{{\bf q}}^2
\\
=
&
-i 4P_{0}^2 g^2
\int_{\epsilon;{\bf k}}
\bar{\Psi}^{\mathrm{q}}(\epsilon;{\bf k}) 
\Psi^{\mathrm{cl}}(\epsilon;{\bf k})
\int_{\bf q} 
\left[ 
{\cal F}^{\mathrm{F}}_{\xi_{{\bf k} + {\bf q}}} 
\frac{2\omega_{\bf q}}{\left( \epsilon - \xi_{{\bf k} + {\bf q}} - i0 \right)^2 - \omega_{\bf q}^2}
+
{\cal F}^{\mathrm{B}}_{\omega_{\bf q}} 
\frac{2\left( \epsilon - \xi_{{\bf k} + {\bf q}}  \right)}{\left( \epsilon - \xi_{{\bf k} + {\bf q}} - i0 \right)^2 - \omega_{\bf q}^2}
\right]
\left(1-\frac{q_{\alpha}^2}{q^2}\right)A_{{\bf q}}^2.
\end{align}
Therefore, we conclude with the expressions for the self-energy, 
\begin{align}
&
\Sigma^{\mathrm{cl-cl}}(\epsilon;{\bf k}) = 0
\\
&
\Sigma^{\mathrm{K}}(\epsilon;{\bf k}) 
= 4P_{0}^2 g^2 (2\pi i)
\int_{\bf q}
\left[
{\cal F}^{\mathrm{F}}_{\xi_{{\bf k} + {\bf q}}} 
{\cal F}^{\mathrm{B}}_{\xi_{{\bf k} + {\bf q}}-\epsilon} 
-1
\right]
\left[
\delta\left( \epsilon - \xi_{{\bf k}+{\bf q}} -\omega_{\bf q}\right) 
-
\delta\left( \epsilon - \xi_{{\bf k}+{\bf q}} +\omega_{\bf q}\right) 
\right]
\left(1-\frac{q_{\alpha}^2}{q^2}\right)A_{{\bf q}}^2,
\\
&
\Sigma^{\mathrm{R}/\mathrm{A}}(\epsilon;{\bf k}) 
= 4P_{0}^2 g^2
\int_{\bf q} 
\left[ 
{\cal F}^{\mathrm{F}}_{\xi_{{\bf k} + {\bf q}}} 
\frac{2\omega_{\bf q}}{\left( \epsilon - \xi_{{\bf k} + {\bf q}} \pm i0 \right)^2 - \omega_{\bf q}^2}
+
{\cal F}^{\mathrm{B}}_{\omega_{\bf q}} 
\frac{2\left( \epsilon - \xi_{{\bf k} + {\bf q}} \right)}{\left( \epsilon - \xi_{{\bf k} + {\bf q}} \pm i0 \right)^2 - \omega_{\bf q}^2}
\right]
\left(1-\frac{q_{\alpha}^2}{q^2}\right)A_{{\bf q}}^2.
\end{align}
Imaginary part of the self-energy reads as
\begin{align}
\mathrm{Im}\Sigma^{\mathrm{R}/\mathrm{A}}(\epsilon;{\bf k}) 
=& \mp 4\pi P_{0}^2 g^2 
\int_{\bf q} 
\left(1-\frac{q_{\alpha}^2}{q^2}\right)
A_{{\bf q}}^2
\\
&
\times
\left\{ 
{\cal F}^{\mathrm{F}}_{\xi_{{\bf k} + {\bf q}}} 
\left[
\delta\left( \epsilon - \xi_{{\bf k}+{\bf q}} -\omega_{\bf q}\right) 
-
\delta\left( \epsilon - \xi_{{\bf k}+{\bf q}} +\omega_{\bf q}\right) 
\right]
+
{\cal F}^{\mathrm{B}}_{\omega_{\bf q}} 
\left[
\delta\left( \epsilon - \xi_{{\bf k}+{\bf q}} -\omega_{\bf q}\right) 
+
\delta\left( \epsilon - \xi_{{\bf k}+{\bf q}} +\omega_{\bf q}\right) 
\right]
\right\}.
\end{align}

Let us now estimate the life-time of electrons.
We use ${\bf e}_{k} = \frac{{\bf k}}{k}$ and ${\bf e}_{0} = \frac{{\bf P}_{0}}{\vert P_{0}\vert}$ unit vectors to write components of the ${\bf q}$ vector. We have
\begin{align}
\frac{q_{\alpha}^2}{q^2} 
= 
\cos^{2}(\theta) ({\bf e}_{k}\cdot {\bf e}_{0})^2 
+ 
\sin^2(\theta)\cos^2(\phi)
\left( \left[ \left[{\bf e}_{k}\times {\bf e}_{0} \right]\times {\bf e}_{k}\right]\cdot{\bf e}_{0}\right)^2
+
2 \cos(\theta)\sin(\theta)\cos(\phi) 
({\bf e}_{k}\cdot {\bf e}_{0})
\left( \left[ \left[{\bf e}_{k}\times {\bf e}_{0} \right]\times {\bf e}_{k}\right]\cdot{\bf e}_{0}\right),
\end{align}
where the third term in the right hand side will not survive the angle integration.
\begin{align}
\cos^2(\theta) = \frac{1}{2kq}\left( Q^2 - k^2 - q^2 \right), ~ \sin^2(\theta) = 1-\cos^{2}(\theta).
\end{align}
\begin{align}
\int_{0}^{2\pi}\frac{d\phi}{2\pi}\cos^{2}(\phi) = \frac{1}{2}.
\end{align}
Then 
\begin{align}
\int\frac{d\phi}{2\pi}\left(1-\frac{q_{\alpha}^2}{q^2}\right)
=
1
-\frac{m}{kq}\left( z - \xi_{k} - \frac{q^2}{2m}\right)({\bf e}_{k}\cdot {\bf e}_{0})^2
-\frac{1}{2}
\left[ 1-\frac{m}{kq}\left( z - \xi_{k} - \frac{q^2}{2m}\right) \right] 
\left( \left[ \left[{\bf e}_{k}\times {\bf e}_{0} \right]\times {\bf e}_{k}\right]\cdot{\bf e}_{0}\right)^2,
\end{align}
where $z=\xi_{Q}$ as before. We pick only the momentum ($z$ and $q$) independent terms from the expression for $\left(1-\frac{q_{\alpha}^2}{q^2}\right)$, and get for the imaginary part of the self-energy 
\begin{align}
\mathrm{Im}\Sigma^{\mathrm{R}/\mathrm{A}}(\epsilon;{\bf k}_{\mathrm{F}}) \approx &
\mp 4\pi P_{0}^2 g^2 \frac{m}{2k}
\int \frac{qdq}{2\pi^2 }A_{{\bf q}}^2
\left[ 1-\frac{1}{2}\left( \left[ \left[{\bf e}_{k}\times {\bf e}_{0} \right]\times {\bf e}_{k}\right]\cdot{\bf e}_{0}\right)^2 \right]
\\
&
\times
  \int_{\xi_{k-q} }^{\xi_{k+q} } dz
  \left\{ 
{\cal F}^{\mathrm{F}}_{z} 
\left[
\delta\left( \epsilon -z -\omega_{\bf q}\right) 
-
\delta\left( \epsilon -z +\omega_{\bf q}\right) 
\right]
+
{\cal F}^{\mathrm{B}}_{\omega_{\bf q}} 
\left[
\delta\left( \epsilon - z -\omega_{\bf q}\right) 
+
\delta\left( \epsilon - z +\omega_{\bf q}\right) 
\right]
\right\}.
\end{align}
The imaginary part of the self-energy is then taken at the mass-shell, i.e. $\epsilon = \xi_{\bf k}$,
\begin{align}
\mathrm{Im}\Sigma^{\mathrm{R}/\mathrm{A}}(\epsilon;{\bf k}_{\mathrm{F}}) \approx &
\mp
 \pi 4P_{0}^2 g^2 \frac{m}{2k_{\mathrm{F}}}
\left[ 1-\frac{1}{2}\left( \left[ \left[{\bf e}_{k}\times {\bf e}_{0} \right]\times {\bf e}_{k}\right]\cdot{\bf e}_{0}\right)^2 \right]
\int_{0}^{2k_{\mathrm{F}}} \frac{qdq}{2\pi^2 }A_{{\bf q}}^2
  \left(
{\cal F}^{\mathrm{F}}_{\epsilon-\omega_{\bf q}} 
-
{\cal F}^{\mathrm{F}}_{\epsilon+\omega_{\bf q}} 
+
2{\cal F}^{\mathrm{B}}_{\omega_{\bf q}} 
\right).
\end{align}
The upper limit is due to the restriction imposed by the delta functions.
Compare this expression with the corresponding one in paragraph 21.3 in Ref. [2]. 
The two expressions are analogous to each other. 
We conclude that the advantage of the Keldysh technique is to skip possible complications occuring in the course of analytic continuation.
We simplify
\begin{align}
1-\frac{1}{2}\left( \left[ \left[{\bf e}_{k}\times {\bf e}_{0} \right]\times {\bf e}_{k}\right]\cdot{\bf e}_{0}\right)^2 
= 1-\frac{1}{2}\sin^2(\phi_{{\bf k}{\bf P}_0}),
\end{align}
where $\phi_{{\bf k}0}$ is an angle between ${\bf k}$ and ${\bf P}_{0}$.

Recalling that $\omega_{\bf q} = \sqrt{\omega_{\mathrm{TO}}^2 + (sq)^2}$ and assuming equilibrium distribution functions, we proceed in estimating the integral. For $\vert \epsilon \vert<\omega_{\mathrm{TO}}$ we get
\begin{align}
\label{lifetime1}
\mathrm{Im}\Sigma^{\mathrm{R}/\mathrm{A}}(\epsilon;{\bf k}_{\mathrm{F}}) \approx &
\pm \pi \frac{8P_{0}^2 g^2}{2\pi^2 } \frac{m}{2k_{\mathrm{F}}}
\frac{\Omega_{0}^2 T}{2\pi s^2}
\left[1-\frac{1}{2}\sin^2(\phi_{{\bf k}{\bf P}_0}) \right]
\\
\times
&
\left\{
\ln\left[ \frac{\cosh\left( \frac{\omega_{\mathrm{TO}} -\epsilon}{2T}  \right) 
\cosh\left( \frac{\omega_{\mathrm{TO}} +\epsilon}{2T}  \right)}{\sinh^2\left( \frac{\omega_{\mathrm{TO}} }{2T}  \right)}\right]
-
\ln\left[ \frac{\cosh\left( \frac{\sqrt{\omega^2_{\mathrm{TO}} + \omega^2_{\mathrm{BG}} }-\epsilon}{2T}  \right) 
\cosh\left( \frac{\sqrt{\omega^2_{\mathrm{TO}} + \omega^2_{\mathrm{BG}} }+\epsilon}{2T}  \right) 
}{\sinh^2\left( \frac{\sqrt{\omega^2_{\mathrm{TO}} + \omega^2_{\mathrm{BG}} } }{2T}  \right)}\right]
\right\}
\\
&
\equiv \pm \left( \frac{P_{0}^2 g^2 m \Omega_{0}^2 \omega_{\mathrm{TO}}}{4\pi^2 s^2 k_{\mathrm{F}}} \right) 
\left[1-\frac{1}{2}\sin^2(\phi_{{\bf k}{\bf P}_0}) \right]
 \frac{2T}{\omega_{\mathrm{TO}}} f(\epsilon, T) ,
\end{align}
where we have made a
\begin{align}
\int \frac{qdq}{\omega_{\bf q}}(..) = \frac{1}{s^2}\int d\omega_{\bf q}(..)
\end{align}
change of variable. 
We assume that $T_{\mathrm{BG}} \gg \omega_{\mathrm{TO}}$.
Here are three limiting behavior of the defined above function for $\epsilon=\xi_{\bf k}=0$, 
\begin{align}
\label{lifetime2}
&
f(0, T) \approx 4 e^{-\frac{\omega_{\mathrm{TO}}}{T}},~~~~ T<\omega_{\mathrm{TO}},
\\
&
\label{lifetime3}
f(0, T) \approx 2 \ln\left(\frac{2T}{\omega_{\mathrm{TO}}} \right),~~~~ T_{\mathrm{BG}}>T>\omega_{\mathrm{TO}},
\\
&
\label{lifetime4}
f(0, T) \approx  2\ln\left(\frac{T_{\mathrm{BG}}}{\omega_{\mathrm{TO}} } \right),~~~~ T>T_{\mathrm{BG}},
\end{align}

%----------------------------------------------------
\subsection{Collision integral. Electric transport.}
%----------------------------------------------------
We anticipate that the resistivity will also have the same temperature dependence as the inverse life-time.
To explicitly check that, we need to construct kinetic equation and solve for the distribution function when the electric field is applied.
Including only the imaginary part of the self-energy, we write for the collision integral
\begin{align}
I_{\mathrm{CI}}(\epsilon;{\bf k}) =
&
i
\Sigma^{\mathrm{K}}(\epsilon;{\bf k}) 
-
i\Sigma^{\mathrm{R}}(\epsilon;{\bf k}) 
{\cal F}_{\epsilon}^{\mathrm{F}}
+
i{\cal F}_{\epsilon}^{\mathrm{F}}
\Sigma^{\mathrm{A}}(\epsilon;{\bf k}) 
=
- \mathrm{Im}\Sigma^{\mathrm{K}}(\epsilon;{\bf k}) 
+ 2{\cal F}_{\epsilon}^{\mathrm{F}}\mathrm{Im} \Sigma^{\mathrm{R}}(\epsilon;{\bf k}) 
\\
=
-
&
(2\pi ) 4P_{0}^2 g^2
\int_{\bf q} 
\left(1-\frac{q_{\alpha}^2}{q^2}\right)
A_{{\bf q}}^2
\left[
\delta\left( \epsilon - \xi_{{\bf k}+{\bf q}} -\omega_{\bf q}\right) 
-
\delta\left( \epsilon - \xi_{{\bf k}+{\bf q}} +\omega_{\bf q}\right) 
\right]
\left({\cal F}^{\mathrm{F}}_{\xi_{{\bf k}+{\bf q}}}{\cal F}^{\mathrm{F}}_{\epsilon} -1  \right)
\\
-
&
(2\pi ) 4P_{0}^2 g^2
\int_{\bf q} 
\left(1-\frac{q_{\alpha}^2}{q^2}\right)
A_{{\bf q}}^2
\left[
\delta\left( \epsilon - \xi_{{\bf k}+{\bf q}} -\omega_{\bf q}\right) 
+
\delta\left( \epsilon - \xi_{{\bf k}+{\bf q}} +\omega_{\bf q}\right) 
\right]
{\cal F}_{\omega_{\bf q}}^{\mathrm{B}}
\left({\cal F}^{\mathrm{F}}_{\epsilon} - {\cal F}^{\mathrm{F}}_{\xi_{{\bf k}+{\bf q}}}  \right).
\end{align}
It can be checked that the collision integral is zero when equilibrium distribution functions with the same temperatures are plugged in.
This can be seen by applying the $\left({\cal F}^{\mathrm{F}}_{\xi_{{\bf k}+{\bf q}}}{\cal F}^{\mathrm{F}}_{\epsilon} -1  \right) = 
-{\cal F}_{\epsilon -\xi_{{\bf k}+{\bf q}} }^{\mathrm{B}}\left( {\cal F}^{\mathrm{F}}_{\epsilon} -  {\cal F}^{\mathrm{F}}_{\xi_{{\bf k}+{\bf q}}} \right)$ identity to the first term, and then substituting $\epsilon -\xi_{{\bf k}+{\bf q}} = \pm \omega_{\bf q}$ there to ${\cal F}_{\epsilon -\xi_{{\bf k}+{\bf q}} }^{\mathrm{B}}$ in accord with corresponding delta-function. 

The collision integral is no longer zero when, for example, the electric field is applied to the system.
In this case we search for the solution in the following form
\begin{align}
&
 {\cal F}^{\mathrm{F}}_{\epsilon} \approx {\cal F}^{\mathrm{F}0}_{\epsilon} + {\cal F}^{\mathrm{F}1}_{\epsilon},
 \\
 &
 {\cal F}_{\omega_{\bf q}}^{\mathrm{B}} = \coth\left( \frac{\omega_{\bf q} }{2T}\right),
\end{align}
where ${\cal F}^{\mathrm{F}0}_{\epsilon} = \tanh\left( \frac{\epsilon}{2T}\right) $ is electron equilibrium distribution function, and to be found ${\cal F}^{\mathrm{F}1}_{\epsilon}$ is proportional to the electric field.
This approximation assumes that both electrons and phonons are at the same temperature.
\begin{align}
&
I_{\mathrm{CI}}(\epsilon;{\bf k}) 
= 
-
8\pi  P_{0}^2 g^2
{\cal F}^{\mathrm{F}1}_{\xi_{\bf k}} 
\\
&
\times
\int_{\bf q} 
\left(1-\frac{q_{\alpha}^2}{q^2}\right)
A_{{\bf q}}^2
\left\{
{\cal F}^{\mathrm{F}0}_{\xi_{{\bf k}+{\bf q}}}
\left[
\delta\left( \epsilon - \xi_{{\bf k}+{\bf q}} -\omega_{\bf q}\right) 
-
\delta\left( \epsilon - \xi_{{\bf k}+{\bf q}} +\omega_{\bf q}\right) 
\right]
+
{\cal F}^{\mathrm{B}}_{\omega_{\bf q}}
\left[
\delta\left( \epsilon - \xi_{{\bf k}+{\bf q}} -\omega_{\bf q}\right) 
+
\delta\left( \epsilon - \xi_{{\bf k}+{\bf q}} +\omega_{\bf q}\right) 
\right]
\right\}
\\
-
&
8\pi  P_{0}^2 g^2
\int_{\bf q} 
\left(1-\frac{q_{\alpha}^2}{q^2}\right)
A_{{\bf q}}^2
\left[
\delta\left( \epsilon - \xi_{{\bf k}+{\bf q}} -\omega_{\bf q}\right) 
-
\delta\left( \epsilon - \xi_{{\bf k}+{\bf q}} +\omega_{\bf q}\right) 
\right]
{\cal F}^{\mathrm{F}0}_{\xi_{{\bf k}}}
{\cal F}^{\mathrm{F}1}_{\xi_{{\bf k}+{\bf q}}}
\\
+
&
8\pi  P_{0}^2 g^2
\int_{\bf q} 
\left(1-\frac{q_{\alpha}^2}{q^2}\right)
A_{{\bf q}}^2
\left[
\delta\left( \epsilon - \xi_{{\bf k}+{\bf q}} -\omega_{\bf q}\right) 
+
\delta\left( \epsilon - \xi_{{\bf k}+{\bf q}} +\omega_{\bf q}\right) 
\right]
{\cal F}^{\mathrm{B}}_{\omega_{\bf q}}
{\cal F}^{\mathrm{F}1}_{\xi_{{\bf k}+{\bf q}}}.
\end{align}
If first integral is proportional to the already calculated imaginary self-energy, the other two require some analysis.
We approximate the collision integral at the mass-shell $\epsilon=\xi_{\bf k}$ and set $\vert k \vert = k_{\mathrm{F}}$, which gives $\xi_{\bf k} = 0$. 
This approximation essentially leaves the dependence of the collision integral only on the angle between the momentum at the Fermi surface and the electric field ${\bf E}$.
Then in the first term out of the remaining two, ${\cal F}^{\mathrm{F}0}_{\xi_{{\bf k}}} = 0$, and the whole term drops out.
In the remaining we assume that we already know the solution, which is
\begin{align}
{\cal F}^{\mathrm{F}1}_{\xi_{{\bf k}+{\bf q}}}\propto \left[({\bf k}+{\bf q})\cdot{\bf E}\right] \delta(\xi_{{\bf k}+{\bf q}}),
\end{align}
then, setting $\epsilon=\xi_{\bf k} = 0$, we get for the expression under the integral
\begin{align}
 \delta(\xi_{{\bf k}+{\bf q}})
 \left[
\delta\left( \epsilon - \xi_{{\bf k}+{\bf q}} -\omega_{\bf q}\right) 
+
\delta\left( \epsilon - \xi_{{\bf k}+{\bf q}} +\omega_{\bf q}\right) 
\right]
\rightarrow
2 \delta(\xi_{{\bf k}+{\bf q}})\delta(\omega_{\bf q}),
\end{align}
which equals zero because $\omega_{\bf q} \neq 0$ away from the transition.
Finally, 
\begin{align}
&
I_{\mathrm{CI}}(\phi_{{\bf kE}}) 
=
- 
8\pi  P_{0}^2 g^2
{\cal F}^{\mathrm{F}1}_{\xi_{\bf k}} 
\\
&
\times
\int_{\bf q} 
\left(1-\frac{q_{\alpha}^2}{q^2}\right)
A_{{\bf q}}^2
\left\{
{\cal F}^{\mathrm{F}0}_{\xi_{{\bf k}+{\bf q}}}
\left[
\delta\left( \epsilon - \xi_{{\bf k}+{\bf q}} -\omega_{\bf q}\right) 
-
\delta\left( \epsilon - \xi_{{\bf k}+{\bf q}} +\omega_{\bf q}\right) 
\right]
+
{\cal F}^{\mathrm{B}}_{\omega_{\bf q}}
\left[
\delta\left( \epsilon - \xi_{{\bf k}+{\bf q}} -\omega_{\bf q}\right) 
+
\delta\left( \epsilon - \xi_{{\bf k}+{\bf q}} +\omega_{\bf q}\right) 
\right]
\right\},
\end{align}
where $\phi_{{\bf kE}}$ is an angle between ${\bf k}$ and ${\bf E}$.
Let us rewrite the collision integral in a more convenient form,
\begin{align}
I_{\mathrm{CI}}(\phi_{{\bf kE}}) = - \frac{1}{\tau_{1{\bf k}}} {\cal F}^{\mathrm{F}1}_{\xi_{\bf k}} .
\end{align}
where we have defined
\begin{align}
\frac{1}{\tau_{1{\bf k}}} = \frac{1}{\tau_{1}}\left[ 1+\cos^2(\phi_{{\bf k}{\bf P}_0})\right]
\end{align}

Expression defining anisotropy of the electric current reads
\begin{align}
\frac{1}{k^2P_{0}^2}
\int_{0}^{2\pi} \frac{d\phi}{2\pi}
\int_{0}^{\pi} \sin(\theta)d\theta {\bf k}
\left( {\bf k}\cdot{\bf E}\right) 
\left( {\bf k}\cdot{\bf P}_{0}\right)^2
= \frac{2}{15} k^2 \left[ {\bf E} + 2({\bf E}\cdot{\bf P}_{0}){\bf P}_{0}\frac{1}{P_{0}^2} \right].
\end{align}

Approximating the kinetic equation in the usual way, we obtain for the electric current
\begin{align}
{\bf j} = e\int_{\bf k}{\bf v}_{\bf k} {\cal F}^{\mathrm{F}}_{\bf k}
&
= 
\frac{2\mu e^2 \tau_{1} \nu_{3\mathrm{D}}}{3m}   {\bf E}
-
\frac{2\mu e^2 \tau_{1} \nu_{3\mathrm{D}}}{15m} 
\left[ {\bf E} + 2({\bf E}\cdot{\bf P}_{0}){\bf P}_{0}\frac{1}{P_{0}^2} \right]
\\
&
=\sigma_{1}{\bf E}- \frac{1}{5}\sigma_{1}\left[ {\bf E} + 2({\bf E}\cdot{\bf P}_{0}){\bf P}_{0}\frac{1}{P_{0}^2} \right],
\end{align}
where $\nu_{3\mathrm{D}} = \frac{m k_{\mathrm{F}}}{2\pi^2}$ is the density of states, 
and $\sigma_{1} = e^2\nu_{3\mathrm{D}}D_{1}$ with $D_{1} = \frac{1}{3}v_{\mathrm{F}}^2 \tau_{1}$ is the conductivity due to the one-phonon scattering processes.
In deriving the expression for the current, we have expanded in $\cos^2(\phi_{{\bf k}{\bf P}_0})$ as
\begin{align}
\frac{1}{ 1+\cos^2(\phi_{{\bf k}{\bf P}_0})} \approx 1-\cos^2(\phi_{{\bf k}{\bf P}_0})
\end{align}
in order to estimate the integrals and extract the main anisotropy of the electric current. 
To complete the analysis, we recall that there are other scattering processes, for example impurity, two-phonon and other scattering processes, in the system which change the collision integral to
\begin{align}
I_{\mathrm{CI}}(\phi_{{\bf kE}}) = 
-\left[ \frac{1}{\tau_{\mathrm{imp}}} + \frac{1}{\tau_{1{\bf k}}}  + \frac{1}{\tau_{2}} \right] 
{\cal F}^{\mathrm{F}1}_{\xi_{\bf k}} ,
\end{align}
where $\frac{1}{\tau_{\mathrm{imp}}}$ and $\frac{1}{\tau_{2}}$ are the inverse scattering times due to the impurities and two-phonons. Both are isotropic in momentum direction. The current will then be
\begin{align}
{\bf j}=\sigma_{\Sigma}{\bf E}- \frac{1}{5}\sigma_{\Sigma}\frac{\tau_{\Sigma}}{\tau_{1}}\left[ {\bf E} + 2({\bf E}\cdot{\bf P}_{0}){\bf P}_{0}\frac{1}{P_{0}^2} \right],
\end{align}
where $\sigma_{\Sigma} =  e^2\nu_{3\mathrm{D}}D_{\Sigma}$ with $D_{1} = \frac{1}{3}v_{\mathrm{F}}^2 \tau_{\Sigma}$ and 
\begin{align}
\frac{1}{\tau_{\Sigma}} =  \frac{1}{\tau_{\mathrm{imp}}} + \frac{1}{\tau_{1}}  + \frac{1}{\tau_{2}}. 
\end{align}
For a general mutual in-plane alignmenent of the ${\bf E}$ and ${\bf P}_{0}$, with an angle $\chi$ between them, the conductivity tensor is
\begin{align}
\hat{\sigma}_{\Sigma} = \left[\begin{array}{cc} \sigma_{xx} & \sigma_{xy} \\ \sigma_{yx} & \sigma_{yy} \end{array}\right],
\end{align}
where 
\begin{align}
\sigma_{xx} = \sigma_{yy} = \sigma_{\Sigma} - \frac{1}{5}\sigma_{\Sigma}\frac{\tau_{\Sigma}}{\tau_{1}}[1+2\cos^2(\chi)], 
~~
\sigma_{xy}=\sigma_{yx} = - \frac{2}{5}\sigma_{\Sigma} \frac{\tau_{\Sigma}}{\tau_{1}}\cos(\chi)\sin(\chi).
\end{align}
Then
\begin{align}
\rho_{xx} = \rho_{yy} = \frac{\sigma_{xx}}{\sigma_{xx}^2 - \sigma_{xy}^2},
~~
\rho_{xy}=\rho_{yx} = -\frac{\sigma_{xy}}{\sigma_{xx}^2 - \sigma_{xy}^2}. 
\end{align}
We can claim that at small temperatures $\rho_{xy} \propto T$ given that $\frac{1}{\tau_{\mathrm{imp}}} > \frac{1}{\tau_{1/2}}$ happening in a range of temperatures $T>\omega_{\mathrm{TO}}$. This is the experimental signature of the ferroelectric metal phase, and is one of the main results of the present paper.

\subsection{References in Supplemental Material}
\begin{enumerate}
\item A. Kamenev, \textit{Field theory of non-equilibrium systems} (Cambridge, University Press, 2012).

\item A.A. Abrikosov, L.P. Gorkov, and I.E. Dzyaloshinskii, \textit{Methods of Quantum Field Theory in Statistical Physics} (Dover Publications, New York, 1975)

\end{enumerate}

\end{widetext}

\end{document}